\documentstyle[12pt]{article}

\begin{document}
\hfill\vbox{
\hbox{CERN-TH/97-337}
\hbox{OHSTPY-HEP-T-97-021}
\hbox{cond-mat/9712041}
\hbox{December 1997} 
\hbox{revised August 1998} 
}\par
\thispagestyle{empty}

\vspace{.5in}

\begin{center}
{\Large \bf Quantum Corrections to the 
  Energy Density of a Homogeneous Bose Gas}

\vspace{.5in}

Eric Braaten$^1$ and Agustin Nieto$^{2}$
\\

$^1$ {\it Department of Physics, The Ohio State University,
Columbus, OH 43210}
\\

$^2$ {\it CERN -- Theory Division, CH-1211 Geneva 23,
        Switzerland}

\end{center}

\vspace{.5in}

\begin{abstract}
Quantum corrections to the properties of a homogeneous interacting Bose
gas at zero temperature can be calculated as a low-density expansion in
powers of $\sqrt{\rho a^3}$, where $\rho$ is the number
density and $a$ is the S-wave scattering length.  We
calculate the ground state energy density to second order in
$\sqrt{\rho a^3}$. The coefficient of the $\rho a^3$
correction
has a logarithmic term that was calculated in 1959. We
present the first calculation of the constant under the
logarithm. The constant
depends not only on $a$, but also on an extra parameter that
describes the low energy $3\to 3$ scattering of the bosons.
In the case of alkali atoms, we argue that the second order
quantum correction is dominated by the logarithmic term,
where the
argument of the logarithm is $\rho a\, \ell_V^2$,
and $\ell_V$ is the length scale set by the van der
Waals potential.

\end{abstract}

\vfill

\newpage

\section{Introduction} 

The successful achievement of Bose-Einstein condensation of
atomic gases in magnetic traps \cite{BEC-Rb,BEC-Li,BEC-Na} has created a
revival of interest in Bose gases.  While a
qualitative description of the condensation can be obtained
using mean field methods \cite{Baym-Pethick}, a more
quantitative treatment requires including corrections from quantum
fluctuations around the mean-field.  The relative magnitude
of these corrections grows as the square root of
the number density of the
atoms.  They will therefore become more important as higher
condensate densities are achieved.

In order to develop a deeper understanding of the quantum
fluctuations, it is worthwhile to go back to the simpler
problem of a homogeneous gas of interacting bosons at zero
temperature.  This problem was studied intensively in the
1950's~\cite{Lee-Yang,Wu}.  The properties of the system can
be calculated as an expansion in powers of $\sqrt{\rho
a^3}$, where $\rho$ is the number density of atoms and $a$
is their S-wave scattering length.  For example, the
expansion for the energy density has the form
\begin{equation}
{\cal E}(\rho) \;=\;  {2 \pi \hbar^2 a \rho^2 \over m}
\left\{ 1 \;+\; {128 \over 15 \sqrt{\pi}} \sqrt{\rho a^3}
        \;+\;  \left[ {8 (4 \pi - 3 \sqrt{3}) \over 3} 
                \ln (\rho a^3) + C \right] \rho a^3
        \;+\; \ldots \right\}  \;.
\label{E-Yang}
\end{equation}
The coefficient of
$\sqrt{\rho a^3}$ was first obtained by Lee and Yang for a
hard sphere gas \cite{Lee-Yang}.  The $\rho a^3$ correction is the 
first term in the expansion that is sensitive to
atomic parameters other than the scattering length.
The coefficient of 
$\ln(\rho)$ in the $\rho a^3$ correction was
calculated by Wu, by Hugenholtz and
Pines, and by Sawada in 1959~\cite{Wu}.

In this paper, we present the first calculation of the
constant $C$ under the logarithm in (\ref{E-Yang}).  We
express the constant in terms of a coupling constant $g_3$
that is defined in terms of the low-energy behavior of the
amplitude for the $3 \to 3$ scattering of atoms in the
vacuum.  The scattering length $a$ and the coupling constant
$g_3$ are the only atomic physics parameters that are needed
to calculate the energy density to second order in the
quantum corrections.  Since $g_3$ is difficult to measure
experimentally or calculate theoretically, it must be
treated as a phenomenological parameter.  The dependence on
this undetermined parameter creates a large uncertainty in
the second order quantum correction, except in cases where
the correction is dominated by the logarithmic term. We
argue that this will typically be the case for alkali atoms,
provided we take the argument of the logarithm to be $\rho
a\, \ell_V^2$, where $\ell_V$ is the length scale set by the
van der Waals potential. In this case our result reduces to
\begin{equation}
{\cal E}(\rho) \;=\;  {2 \pi \hbar^2 a \rho^2 \over m}
\left\{ 1 \;+\; {128 \over 15 \sqrt{\pi}} \sqrt{\rho a^3}
        \;+\;  {8 (4 \pi - 3 \sqrt{3}) \over 3} 
                \ln \left(\rho a\, \ell_V^2 \right)\rho a^3
        \;+\; \ldots \right\}  \;.
\label{E-lsl}
\end{equation}
If the logarithm is large compared to 1, this should give an
accurate estimate of the second order quantum corrections 
to the energy density.

We begin in Section~\ref{QFT} by formulating the problem of
calculating the energy density as a quantum field theory
problem.  In Section~\ref{EdensityinD}, we set up a
perturbative framework and use it to calculate the energy
density to second order in the quantum corrections.  In
Section~\ref{scattering}, we calculate the second order
quantum correction to the $T$-matrix element for the $3 \to
3$ scattering of atoms in the vacuum.  We define the
coupling constant $g_3$ and show that the renormalization of
$g_3$ is necessary to remove a logarithmic ultraviolet
divergence from the $T$-matrix element.  In
Section~\ref{renormalization}, we complete the calculation
of the energy density by including the contribution from
$g_3$ and its renormalization.  We then discuss the case of
alkali atoms, and argue that the dependence on $g_3$ can be
eliminated in favor of a logarithmic dependence on the
length scale set by the van der Waals interaction.  The
two-loop Feynman diagrams that contribute to the energy
density are calculated in the Appendix.

\section{Field Theory Formulation}
\label{QFT}

We are interested in computing the ground state energy 
density ${\cal E}$ of a homogeneous Bose gas 
as a function of its density $\rho$. 
Our starting point is a local quantum field theory that
describes atoms with momenta much lower than the inverse of the
range of the interatomic potential, which is several \AA \
in the case of alkali atoms. 
At such low energies, the interactions
appear pointlike on the scale of the de Broglie wavelengths
of the atoms.  The many-body quantum mechanics of the atoms can 
therefore be formulated in terms of a quantum
field theory whose hamiltonian density is a local
function of the field:
\begin{equation}
{\cal H} \;=\;
{\hbar^2 \over 2m} \mbox{\boldmath $\nabla$} \psi^\dagger  
        \cdot \mbox{\boldmath $\nabla$}\psi
\;+\; {1 \over 4} g\, (\psi^\dagger \psi)^2
\;+\; {1 \over 36} g_3\, (\psi^\dagger \psi)^3
\;+\; \cdots .
\label{eq:hami}
\end{equation}
For simplicity, we have assumed that the atoms have only 
one spin state
so that they can be represented by a single complex field
$\psi({\bf x},t)$.  The $(\psi^\dagger \psi)^2$ term
represents $2\to 2$ scattering through an S-wave interaction
with scattering length $a$ given by 
\begin{equation}
g \;=\; {8 \pi \hbar^2 a \over m}.
\label{g-a}
\end{equation}
This coupling constant contains all the information about atomic
interactions that is required to calculate the first order 
quantum corrections to the properties of a sufficiently cold and 
dilute Bose gas. 
We follow the effective-field-theory philosophy~\cite{Georgi}
of including in the hamiltonian all possible local terms that are 
consistent with the symmetries, 
which include Galilean invariance 
and the phase symmetry $\psi\rightarrow e^{i\alpha} \psi$.
The term $(\psi^\dagger \psi)^3$ in (\ref{eq:hami}) allows 
$3 \to 3$ scattering through a pointlike interaction.
The $\ldots$'s in (\ref{eq:hami}) include all possible terms that are
higher order in the derivatives or in the number of fields.
In principle, the
coefficients of these terms can be calculated from the
$n$-body potentials that describe interatomic interactions.
In the absence of such calculations, they can be taken as
phenomenological parameters.
The effective-field-theory philosophy is based on the assumption that
there is a systematic expansion in powers of the momentum. 
The relative importance of the interactions terms 
in (\ref{eq:hami}) at a low momentum scale $p$ is then given by
the dimensionless combination of the coupling constant and $p$,
which is $m g p$ and $m g_3 p^4$ for the $(\psi^\dagger \psi)^2$ term 
and the $(\psi^\dagger \psi)^3$ term, respectively.
If $p$ is sufficiently small, the effects of the $g_3$ term will
be much smaller than those of the $g$ term.
Terms with more derivatives or with higher powers of
$\psi$ give effects that are suppressed by even more powers of $p$.
Of the infinitely many terms in (\ref{eq:hami}),
there are only a finite number that
contribute at any given order in $p$.
In the case of the energy density, the appropriate 
momentum scales are $\rho^{1/3}$ and $(\rho a)^{1/2}$,
so the momentum expansion becomes an expansion in powers of the density.
We will find that
the scattering length $g$ and the coupling constant $g_3$ 
are the only atomic physics parameters 
that contribute to the energy density through
third order in $\rho$~\cite{Braaten-Nieto}. 

At a given order in $p$, only a finite number of terms contribute.
By tuning the coefficients of these terms, one can
describe $n \to n$ scattering of atoms in the vacuum with
whatever accuracy is desired.

The phase symmetry $\psi\rightarrow e^{i\alpha} \psi$ 
of the hamiltonian
implies the conservation of the number of atoms. The number density
operator is ${\cal N} = \psi^\dagger\psi$.
A homogeneous Bose gas can be described by a field theory
with hamiltonian density ${\cal H}-\mu\,{\cal N}$, where
$\mu$ is the chemical potential. The energy density ${\cal
E}(\mu)$ and the number density $\rho(\mu)$ in the ground
state of this field theory are
\begin{eqnarray}
  \label{Eofmu}
  {\cal E}(\mu) &= &\langle{\cal H}\rangle_\mu \,,
\\
  \label{rhofmu}
  \rho(\mu) &= &\langle{\cal N}\rangle_\mu \,,
\end{eqnarray}
where $\langle\cdots\rangle_\mu$ denotes the expectation
value in the ground state. By eliminating $\mu$ from the two
equations~(\ref{Eofmu}) and~(\ref{rhofmu}), we obtain ${\cal
E}$ as a function of $\rho$. It is simpler in perturbative
calculations to first calculate the free energy density
${\cal F}(\mu)$ of the ground state:
\begin{equation}
  {\cal F}(\mu) \equiv \langle{\cal H} - 
    \mu\,{\cal N}\rangle_\mu \,.
\end{equation}
After inverting~(\ref{rhofmu}) to obtain $\mu$ as a function
of $\rho$, one can obtain the energy density from
\begin{equation}
{\cal E} \;=\; {\cal F} \;+\; \mu \rho \,.
\label{E-rho}
\end{equation}

The partition function for the field theory with hamiltonian
density ${\cal H}-\mu\,{\cal N}$ can be expressed as a functional
integral:
\begin{equation}\label{parfun}
  {\cal Z} = \int\,{\cal D}\psi^\dagger\,{\cal D}\psi\,
        \exp \left\{i\,S[\psi]\right\}  \,,  
\end{equation}
where the action $S[\psi]$ is given by
\begin{equation}
S[\psi] \;=\; 
\int dt\,\int d^3x\;
\left\{ \psi^\dagger 
        \left( i\partial_t + {\mbox{\boldmath
                        $\nabla$}^2\over 2m} + \mu \right) \psi
        - {1 \over 4} g\, (\psi^\dagger \psi)^2
        - {1 \over 36} g_3\, (\psi^\dagger \psi)^3
        - \ldots\right\} \,.  
\label{actpsi}
\end{equation}
We have set $\hbar = 1$ in the action.  Dimensional analysis
can be used to reinsert the factors of $\hbar$ at the end of
the calculation. The $\ldots$'s in~(\ref{actpsi}) represents
all possible
terms with higher powers of $\psi$ or more factors of $\mbox{\boldmath
$\nabla$}$.
They also include counterterms that are needed to cancel ultraviolet
divergences associated with the parameters $\mu$, $g$, and
$g_3$. 
The free energy density ${\cal F}(\mu)$ is related to
the partition function by
\begin{equation}\label{fene}
  {\cal Z} \;=\; \exp{\left\{-iVT\,{\cal F}(\mu)\right\} } \,,
\end{equation}
where $VT$ is the spacetime volume. The ground state expectation 
value of an operator $\langle{\cal O}\rangle$ can be expressed as 
a functional integral:
\begin{equation}\label{evO}
   \langle{\cal O}\rangle_\mu \;=\; {1\over{\cal Z}} \int {\cal D}\psi^\dagger
      {\cal D}\psi\, {\cal O}\exp \{i S[\psi] \} \,.
\end{equation}
By differentiating the logarithm of both~(\ref{parfun}) and~(\ref{fene}) with
respect to $\mu$ and using~(\ref{evO}), we obtain the relation
\begin{equation}\label{rho-mu}
      \rho(\mu) \;=\; - {d{\cal F}\over d\mu}(\mu)\,.
   \end{equation}
Differentiating~(\ref{E-rho}) with respect to $\rho$ and
using~(\ref{rho-mu}), we obtain
\begin{equation}\label{dedrho}
      {d{\cal E}\over d\rho}(\rho) = \mu(\rho)\,.
   \end{equation}
The simplest way to calculate the energy density is to first
calculate ${\cal F}(\mu)$, use~(\ref{rho-mu}) to get 
$\rho(\mu)$, invert to get
$\mu(\rho)$, and then integrate~(\ref{dedrho}) to get ${\cal
E}(\rho)$.

It is convenient to parameterize the quantum field
$\psi({\bf r},t)$ in terms of two real-valued quantum
fields $\xi$ and $\eta$ that describe quantum fluctuations
around an arbitrary constant background $v$:
\begin{equation}
  \psi({\bf r},t) \;=\;  v
  \;+\; { \xi({\bf r},t) + i \eta({\bf r},t) \over \sqrt{2} } \,.
\label{psi-cart}
\end{equation}
After inserting the field parameterization~(\ref{psi-cart}) into
the action (\ref{actpsi}), it
can be expanded in powers of the
quantum fields $\xi$ and $\eta$.
By separating the action into a free part and an interaction 
part, we can express the thermodynamic functions as diagrammatic
expansions.
The free energy density ${\cal F}$ is the
sum of all {\em connected  vacuum diagrams}, which are Feynman
diagrams with no external legs. This sum is independent
of the arbitrary background $v$. It is
convenient to define the thermodynamic potential
$\Omega(\mu,v)$, which is the sum of all
{\it one-particle-irreducible} vacuum diagrams. The
thermodynamic potential, which depends on $v$, contains
the information required to determine all of the thermodynamic
functions. The free energy ${\cal F}(\mu)$ can be obtained by
evaluating $\Omega(\mu,v)$
at a particular value of $v$ given by the {\em tadpole
condition\/}
\begin{equation}\label{tadpole-cond}
      \overline{v}(\mu) = \langle\psi\rangle_\mu  \,.
   \end{equation}
For this value of $v$, those diagrams that can be
disconnected by cutting a single line vanish. Thus the sum 
of connected vacuum diagram reduces to the sum of 
one-particle-irreducible vacuum diagrams and
we have
\begin{equation} \label{F-Omega}
      {\cal F}(\mu) \;=\; \Omega(\mu,\overline{v}(\mu)) \,.
   \end{equation}
Using~(\ref{psi-cart}), the tadpole
condition~(\ref{tadpole-cond}) reduces to
$\langle\xi\rangle_\mu=\langle\eta\rangle_\mu=0$. The phase of the
field $\psi$ can be chosen so that $\langle\eta\rangle_\mu$ is
automatically 0. The condition $\langle\xi\rangle_\mu=0$ can be
conveniently expressed in terms of the thermodynamic
potential itself:
\begin{equation} \label{Omega-min}
     {\partial\Omega\over\partial v}(\mu,\overline{v}(\mu)) \;=\; 0 \,.
   \end{equation}
Differentiating both sides of~(\ref{F-Omega}) with respect
to $\mu$ and
using~(\ref{Omega-min}), we obtain
\begin{equation}
      {d{\cal F}\over d\mu}(\mu) \;=\;
         {\partial\Omega\over\partial\mu}(\mu,\overline{v}(\mu)) \,.
   \end{equation}
Comparing with~(\ref{rho-mu}), we find that the number density
can be expressed as
\begin{equation}
      \rho(\mu) \;=\; 
        - {\partial\Omega\over\partial\mu}(\mu,\overline{v}(\mu)) \,.
   \end{equation}

\section{Ground State Energy Density}
\label{EdensityinD}

In this section, we calculate the ground state energy for a 
homogeneous Bose gas
to second order in the quantum corrections. 
We first set up a perturbative framework for carrying out 
calculations in the presence of a nonzero chemical potential.
We use the framework to calculate the energy density to second order in
the quantum corrections.  We then carry out the renormalizations of 
$\mu$ and $g$ that are necessary to remove 
power ultraviolet divergences from the energy density.

\subsection{Perturbative framework}

We can describe a Bose gas with nonzero density
$\rho$ by the action (\ref{actpsi})
with an appropriately chosen value 
of the chemical potential. 
For simplicity, we set $g_3=0$ 
and omit all terms in~(\ref{actpsi}) that are 
higher order in $\psi$ or $\mbox{\boldmath $\nabla$}$. 
We ignore for the moment the counterterms associated with
renormalization, so the parameters $\mu$ and $g$ should 
be regarded as bare parameters. Inserting the field 
parameterization~(\ref{psi-cart}) into the action and expanding
in powers of $\xi$ and $\eta$, the action becomes
\begin{eqnarray}
  S[\psi] &=&   \int dt \int d^3x \Bigg\{ 
    \mu v^2 - {1\over 4} g v^4 
    + {v X\over \sqrt{2} m} \xi
    \;+\; {1 \over 2} \left( \eta \dot{\xi} - \xi \dot{\eta} \right)
    \;+\; {1 \over 4m} \xi \left( \mbox{\boldmath $\nabla$}^2 
        - 2mgv^2 + X \right) \xi
\nonumber \\ && \qquad\qquad
    \;+\; {1 \over 4m} \eta \left( \mbox{\boldmath $\nabla$}^2 
        + X \right) \eta
    \;-\;  {g v \over \sqrt{8}} \xi \left( \xi^2 + \eta^2 \right) 
    \;-\;  {g \over 16} \left( \xi^2 + \eta^2 \right)^2
  \Bigg\}\,,
\label{S-cart}
\end{eqnarray}
where $\dot{f} \equiv {\partial \ \over \partial t} f$ and 
\begin{equation}
      X = 2m\left( \mu - {1\over 2} g v^2 \right) \,.
   \label{X-def}
   \end{equation}

To organize the quantum corrections into a loop expansion,
we separate the terms in the action~(\ref{S-cart}) that
depend on $\xi$ and $\eta$ into a free part and an
interaction part:
\begin{equation}
S[\psi] \;=\;  S[v] 
\;+\; S_{\rm free}[\xi,\eta] \;+\; S_{\rm int}[v,\xi,\eta]  \,.
\label{S-decomp}
\end{equation}
The free part of the action consists of the terms quadratic in 
$\xi$ and $\eta$: 
\begin{eqnarray}
S_{\rm free}[\xi,\eta] &=&  \int dt \int d^3x \Bigg\{ 
{1 \over 2} \left( \eta \dot{\xi} - \xi \dot{\eta} \right)
\nonumber \\
&&\qquad
\;+\; {1 \over 4 m} \xi ( \mbox{\boldmath $\nabla$}^2 
        - 2 mgv^2 + X ) \xi
\;+\; {1 \over 4 m} \eta ( \mbox{\boldmath $\nabla$}^2 
        + X ) \eta \Bigg\} \,.
\label{S-free}
\end{eqnarray}
The Fourier transform of the propagator for the fields 
$\xi$ and $\eta$ is a $2 \times 2$ matrix:
\begin{equation}
D(\omega,k,v) = 
{i \over  \omega^2 - \varepsilon^2(k,v) + i\epsilon}
\left( 
\begin{array}{cc}
        (k^2 - X)/2m & - i \omega       \\
        i \omega    & (k^2 + 2mgv^2 - X)/2m 
        \end{array}
\right) \,,
\label{propagator}
\end{equation}
where ${\bf k}$ is the wavevector, $\omega$ is the frequency, and
\begin{equation}
  \varepsilon^2(k,v) = {1\over 4m^2} 
    (k^2 + 2mgv^2 - X)(k^2 - X) \,.
\label{energy-k}
\end{equation}
The diagonal elements of the propagator matrix
(\ref{propagator}) are represented by solid lines for $\xi$
and dashed lines for $\eta$, as illustrated in Figs.~1(a)
and~1(b).  The off-diagonal elements are represented by a
line that is half solid and half dashed, as in Fig.~1(c).
All the remaining terms in the action (\ref{S-cart}) are
treated as interactions:
\begin{equation}
S_{\rm int}[v,\xi,\eta] \;=\; \int dt \int d^3x \Bigg\{ 
{v X\over \sqrt{2} m} \xi
\;-\;  {gv \over \sqrt{8}} \xi \left( \xi^2 + \eta^2 \right) 
\;-\;  {g \over 16} \left( \xi^2 + \eta^2 \right)^2 
\Bigg\} \,.
\label{Sint-cart}
\end{equation}
The term proportional to $\xi$ is represented by a dot at
which a solid line terminates as illustrated in
Fig.~1(d). The 3 and 4-point couplings are represented
by points that connect three and four lines, respectively,
as in Figs.~1(e)--1(i).

It is possible to diagonalize the propagator matrix
(\ref{propagator}) by applying a Bogoliubov transformation
to the fields $\xi$ and $\eta$. However,
such a transformation makes the interaction terms  in the
action significantly more complicated and increases the
number of diagrams that contribute to most quantities. For
explicit calculations, it is more economical to  minimize
the number of diagrams.  We therefore prefer to use a 
propagator matrix with off-diagonal elements. This
perturbative framework was recently used by Haugset,
Haugerud, and Ravndal~\cite{ravndal} to reproduce the
$\sqrt{\rho a^3}$ correction in the
expression~(\ref{E-Yang}) for the energy density.

\subsection{Free energy density}

If the $n$-loop contribution to the thermodynamic potential
$\Omega$ is
denoted by $\Omega_n(\mu,v)$, the loop expansion
for the free energy density (\ref{F-Omega}) is
\begin{equation}\label{F-loop}
      {\cal F}(\mu) \;=\; \Omega_0(\mu,\overline{v}) +
         \Omega_1(\mu,\overline{v}) \;+\;
         \Omega_2(\mu,\overline{v}) \;+\; \cdots \,,
   \end{equation}
where $\overline{v}$ is the condensate, which satisfies (\ref{Omega-min}):
\begin{equation}\label{cond-eq}
      {\partial\Omega_0\over\partial v}(\mu,\overline{v}) \;+\;
         {\partial\Omega_1\over\partial v}(\mu,\overline{v}) \;+\;
         \cdots \;=\; 0 \,.
   \end{equation}
The loop expansion~(\ref{F-loop}) does not coincide with the
expansion  in the order of quantum corrections because of its 
dependence on $\overline{v}$. To obtain the quantum expansion,
we must expand the condensate $\overline{v}$ around its
classical value $\overline{v}_0$, which satisfies
\begin{equation}\label{vev0}
      {\partial\Omega_0\over\partial v}(\mu,\overline{v}_0) \;=\; 0 \,.
   \end{equation}
By expanding~(\ref{cond-eq}) in powers of
$\overline{v}-\overline{v}_0$, and solving for $\overline{v}$, we
obtain the quantum expansion for the condensate:
\begin{equation}\label{cond-loop}
      \overline{v}(\mu) \;=\; \overline{v}_0(\mu) \;+\; 
         \overline{v}_1(\mu) \;+\;
         \overline{v}_2(\mu) \;+\; \cdots \,,
   \end{equation}
where $\overline{v}_n$ is the $n$-th order quantum correction. For
example, the first-order quantum correction is
\begin{equation}\label{31}
      \overline{v}_1(\mu) \;=\; -{\partial\Omega_1\over\partial v}
            (\mu,\overline{v}_0)
         \left/{\partial^2\Omega_0\over\partial v^2}
            (\mu,\overline{v}_0)\right.\,.
   \end{equation}
Inserting~(\ref{cond-loop}) into~(\ref{F-loop}) and expanding in powers
of $\overline{v}_1$, $\overline{v}_2$, $\ldots$, we obtain the quantum 
expansion for the free energy. Keeping only terms through second order,
we have
\begin{equation}\label{free}
      {\cal F}(\mu) = \Omega_0(\mu,\overline{v}_0) + \Omega_1(
         \mu,\overline{v}_0) + 
         \left(\Omega_2(\mu,\overline{v}_0) +
         \overline{v}_1
            {\partial\Omega_1\over\partial v}
            (\mu,\overline{v}_0)+
         {1\over 2}\overline{v}_1^2 \,
            {\partial^2\Omega_0\over\partial v^2}(\mu,\overline{v}_0)
      \right) \,.
   \end{equation}

The mean-field contribution to $\Omega(\mu,v)$ is given by
the terms in~(\ref{S-cart}) that are independent of $\xi$
and $\eta$:
\begin{equation}\label{ome0}
  \Omega_0(\mu,v) = - \mu v^2 + {1\over 4} g v^4 \, .
\end{equation}
One solution to (\ref{vev0}) is $\overline{v}_0=0$, but it
is a local maximum of $\Omega_0$ and therefore represents an
unstable configuration. The stable solution is
\begin{equation}\label{v0}
  \overline{v}_0^2 = 2\mu/g \, .
\end{equation}
The tree-level contribution to 
${\cal F}(\mu)$ given in~(\ref{free}) is
\begin{equation}  \label{ome0-v0}
  \Omega_0(\mu,\overline{v}_0) = -{\mu^2\over g} \, . 
\end{equation}
The dispersion relation~(\ref{energy-k})
simplifies significantly at the point $v=\overline{v}_0$,
because $X=0$ at that point.  It reduces to the Bogoliubov
dispersion relation:
\begin{equation}
\varepsilon(k) \;\equiv\; 
        \varepsilon(k,\overline{v}_0) \;=\; 
        {k\sqrt{k^2+\Lambda^2}\over 2m} \,,
\label{Bogol}
\end{equation}
where $\Lambda^2=4m\mu$. The tadpole interaction in~(\ref{Sint-cart})
also vanishes when $v=\overline{v}_0$.

Using the free part of the action~(\ref{S-free}), 
we can obtain the one-loop contribution to
$\Omega(\mu,v)$:
\begin{equation}
      \Omega_1(\mu,v) = {i\over 2}
      \int {d^3 k\over (2\pi)^3}
      \int {d\omega\over 2\pi} \ln\det D(\omega,k,v) \,,
\end{equation}
where $D(\omega,k,v)$ is given in~(\ref{propagator}). By
integrating over $\omega$, we obtain
\begin{equation}\label{ome1}
  \Omega_1(\mu,v) = 
        {1\over 2}\int {d^3 k\over (2\pi)^3}\,\varepsilon(k,v) \, ,
\end{equation}
where $\varepsilon(k,v)$ is given by~(\ref{energy-k}). 
The one-loop contribution to the free energy is
\begin{equation} \label{ome1-v0}
\Omega_1(\mu,\overline{v}_0) \;=\; 
        {1\over 4m} I_{0,-1}(4m\mu) \, ,
\end{equation}
where $I_{0,-1}$ is a function of $\Lambda^2=4m\mu$ defined
by the integral
\begin{equation}
I_{m,n}(\Lambda^2) \;=\; \int {d^3 p\over (2\pi)^3}
         {(p^2)^m\over [2m\varepsilon(p)]^n} \, .
\label{I-mn}
\end{equation}
Differentiating the expression~(\ref{ome1}) with respect to
$v$ and evaluating at $v=\overline{v}_0$, we obtain
\begin{equation}\label{ome2a}
{\partial\Omega_1\over\partial v}(\mu,\overline{v}_0) \;=\; 
      {g\overline{v}_0 \over 4} ( 3I_{1,1} + I_{-1,-1} ) \,.
\end{equation}
The first order quantum correction to the condensate, which
is given by~(\ref{31}) is
\begin{equation}\label{41.5}
\overline{v}_1 \;=\; 
- {g \overline{v}_0 \over 16 \mu} (3 I_{1,1} + I_{-1,-1})\, .
\end{equation}

The two-loop contribution to the thermodynamic potential
is obtained from the vacuum diagrams shown in Fig.~\ref{twoloop}. 
The contributions from the individual diagrams 
are given in Appendix~\ref{twoloopdia}.
The sum of the diagrams gives
\begin{eqnarray} \label{ome2b}
\Omega_2(\mu,\overline{v}_0) \;=\; 
{m g \mu \over 8} \, J
\;+\; {g \over 64}\, \left[ 
      3 I_{-1,-1}^2 + 2 I_{-1,-1} I_{1,1} + 3 I_{1,1}^2 \right] \, ,
\end{eqnarray}
where
\begin{equation}
J \;=\; 6 J_{0, 0, 1} - J_{-1, -1, 1} 
        - 3 J_{1, 1, 1}  - 2 J_{-1, 0, 0} \, .
\label{J-def}
\end{equation}
The integrals $J_{l,m,n}$ are functions of $\Lambda^2=4m\mu$ defined by
\begin{equation}
J_{l,m,n}(\Lambda^2) = \int {d^3 p\over (2\pi)^3}
         \int {d^3 q\over (2\pi)^3}{
         \left[p^2/ 2m\varepsilon(p)\right]^l
         \left[q^2/ 2m\varepsilon(q)\right]^m
         \left[r^2/ 2m\varepsilon(r)\right]^n
         \over 2m[ \varepsilon(p)+\varepsilon(q)+\varepsilon(r) ]} \,,
\label{J-lmn}
\end{equation} 
where $r=|{\bf p}+{\bf q}|$.

Inserting~(\ref{ome0-v0}), (\ref{ome1-v0}), (\ref{ome2a}),
(\ref{41.5}), and~(\ref{ome2b}) into~(\ref{free}), we
obtain the complete  expression for the  free energy
density to second order  in the quantum corrections:
\begin{equation}\label{Fmu}
{\cal F}(\mu) = -{\mu^2\over g}   
    + {1\over 4m}I_{0,-1} (4m\mu)
    + {mg\mu\over 8} J
    + {g\over 32} \left[ 
      I_{-1,-1}^2 - 2 I_{-1,-1} I_{1,1} - 3 I_{1,1}^2 \right]
     \,.
\end{equation}
The free energy~(\ref{Fmu}) depends on $\mu$ through the explicit
factors of $\mu$ and through the momentum scale of the integrals
which is $\Lambda^2=4m\mu$. We have made the argument of
the integral explicit for the $I_{0,-1}$ term
in~(\ref{Fmu}).

\subsection{Energy density}
\label{energy-density}

To calculate the energy density, we use~(\ref{rho-mu}) to get
$\rho(\mu)$, invert that relation to get $\mu(\rho)$, and
integrate using~(\ref{dedrho}) to get ${\cal E}(\rho)$.
Differentiating~(\ref{Fmu}) and using~(\ref{rho-mu}), the
number density is
\begin{equation}\label{rhomu}
      \rho(\mu) = {2\mu\over g} - {1\over 2}
      I_{1,1}(4m\mu)  
      -{mg\over 8}
      \left[
         J + 4 m \mu J'
         + (I_{-1,-1}-I_{1,1}) I_{0,1}
         + (I_{-1,-1}+3I_{1,1}) I_{2,3}
      \right]\,,
\end{equation}
where $J'=dJ/d\Lambda^2$ and all the integrals are
functions of $\Lambda^2=4m\mu$. We have made the argument
of the integral explicit for the $I_{1,1}$
term. We have
used the identity~(\ref{prop1}) to differentiate the
integrals $I_{m,n}$. Inverting~(\ref{rhomu}) to get $\mu$
as a function of $\rho$, we obtain
\begin{equation} \label{murho}
      \mu(\rho) = {1\over 2}g\rho + {1\over 4} g
      I_{1,1}(2mg\rho)
      + {mg^2\over 16}
      \left[
         J + 2mg\rho J'
         + (I_{-1,-1}-I_{1,1}) I_{0,1}
         + (I_{-1,-1}+I_{1,1}) I_{2,3}
      \right]\,,
   \end{equation}
where the integrals are now functions of
$\Lambda^2=2mg\rho$. Using the identity~(\ref{prop1}), we
can write the expression as a total derivative:
\begin{equation}\label{murhod}
\mu(\rho) \;=\; {d\ \over d\rho}
      \left\{
         {1\over 4} g\rho^2 + {1\over 4m} I_{0,-1}(2mg\rho)
         + {mg^2\rho\over 16}J
         + {g\over 32}
         \left[
            I_{-1,-1}^2 -2I_{-1,-1}I_{1,1}-I_{1,1}^2
         \right]
      \right\} \,.
   \end{equation}
We can now read off the energy density using~(\ref{dedrho}):
\begin{equation} \label{Erho} 
      {\cal E}(\rho) = {\cal E}_0 + {1\over 4}g\rho^2
      + {1\over 4m}I_{0,-1}(2mg\rho)
      + {mg^2\rho\over 16}J
      +{g\over 32}
      \left[ I_{-1,-1}^2-2I_{-1,-1}I_{1,1}-I_{1,1}^2 \right] \,,
   \end{equation}
where ${\cal E}_0$ is an integration constant and
all the integrals are functions of $\Lambda^2=2mg\rho$.
It is convenient to choose the integration constant ${\cal
E}_0$ so that the energy of the vacuum is zero: ${\cal
E}(0)=0$.

\subsection{Renormalization of $\mu$ and $g$}

Our result (\ref{Erho}) for the energy density can be 
generalized to an arbitrary number of spatial dimensions $D$ simply by 
replacing the integration measure $d^3p / (2 \pi)^3$ in 
(\ref{I-mn}) and (\ref{J-lmn}) by $d^Dp / (2 \pi)^D$.
The integrals $I_{m,n}$ and $J_{l,m,n}$ in (\ref{Erho}) are ultraviolet
divergent for any positive number of dimensions $D$. 
If we impose a momentum cutoff $\Lambda_{UV}$, then $I_{0,-1}$
diverges like $\Lambda_{UV}^{D+2}$, while $I_{-1,-1}$ and
$I_{1,1}$ diverge like $\Lambda_{UV}^D$. The integrals
$J_{l,m,n}$ contain subintegrals that diverge like
$\Lambda_{UV}^{D-2}$, and, if $D>1$, they also have 
an overall
divergence that scales like $\Lambda_{UV}^{2D-2}$. There are
cancellations among the $J_{l,m,n}$ integrals that reduce
the overall divergence to $\Lambda_{UV}^{2D-6}$ for $D>3$ and
$\ln\Lambda_{UV}$ for $D=3$.

The divergences can be removed by renormalization. A
convenient way to implement the renormalization is to add
counterterms to the action~(\ref{actpsi}): 
\begin{equation}
\delta S \;=\; \int dt \int d^3x 
      \left\{ 
         \delta\mu\,\psi^\dagger\psi 
         -{1\over 4}\delta g (\psi^\dagger\psi)^2 
         +\cdots
      \right\}\,. 
\end{equation} 
In perturbative calculations, the counterterms $\delta\mu$
and $\delta g$ should be treated as quantum corrections.
They can be expanded according to the order in the quantum
correction:
\begin{eqnarray}
      \delta\mu &=& \delta_1\mu + \delta_2\mu + \cdots\,,
   \\
      \delta g &=& \delta_1 g + \delta_2 g + \cdots\,.
\end{eqnarray}
To obtain the free energy after the renormalizations of
$\mu$ and $g$, we substitute
$\mu\to\mu+\delta\mu$ and $g\to g+\delta g$
into~(\ref{Fmu}) and expand in the
order of the quantum correction. The complete expression to
second order in the quantum corrections is
\begin{eqnarray}
      {\cal F}(\mu) &=& -{\mu^2\over g} + 
      \left[
         {1\over 4m} I_{0,-1}(4m\mu) 
         - 2 {\mu\over g}\delta_1\mu 
         + {\mu^2\over g^2}\delta_1 g
      \right]
   \nonumber \\ &&
      +
      \left[
         {mg\mu\over 8}J 
         + {g\over 32}
            (I_{-1,-1}^2-2I_{-1,-1}I_{1,1}-3I_{1,1}^2)
         + {1\over 2} I_{1,1}\delta_1\mu
      \right.
   \nonumber \\ &&
      \left. \qquad \qquad \qquad
         - 2{\mu\over g}\delta_2\mu + {\mu^2\over g^2}\delta_2 g
         - {1\over g}(\delta_1\mu)^2 
         + 2{\mu\over g^2}\delta_1\mu\,\delta_1 g 
         - {\mu^2\over g^3}(\delta_1 g)^2
      \right] \,.
\label{F-mu-ren}
\end{eqnarray}
By repeating each of the
steps in Section~\ref{energy-density} including the effects
of the counterterms, we obtain an 
expression for the energy density that takes into account
the renormalization of $\mu$ and $g$:
\begin{eqnarray}
{\cal E}(\rho) &=& {\cal E}_0 +{1\over 4} g\rho^2 + 
\left[ {1\over 4m} I_{0,-1}(2mg\rho) 
         - \rho\,\delta_1\mu 
         + {\rho^2\over 4}\delta_1 g
      \right]
\nonumber \\ &&
\;+\; \left[
         {mg^2\rho\over 16}J 
         + {g\over 32} (I_{-1,-1}^2-2I_{-1,-1}I_{1,1}-I_{1,1}^2)
         + {\rho\over 4} I_{1,1}\delta_1 g
         - \rho\,\delta_2\mu 
         + {\rho^2\over 4}\delta_2 g
      \right] \,.
\label{E-rho-ren}
\end{eqnarray}
The ultraviolet divergences in~(\ref{E-rho-ren}) that are independent 
of $\rho$ can be 
cancelled by ${\cal E}_0$.
The counterterms $\delta_1 \mu$, $\delta_2 \mu$, $\delta_1 g$, 
and $\delta_2 g$
can be determined by demanding that the $\rho$-dependent 
power ultraviolet divergences cancel. 
For example, the integral appearing in the first
order quantum correction can be written
\begin{eqnarray}
I_{0,-1} &=& \int{d^3p \over (2\pi)^3}
\left( p \sqrt{p^2 + 2 m g \rho} - p^2 - m g \rho +
            {m^2 g^2 \rho^2 \over 2 p^2} 
            \right)
\nonumber \\ 
&& \;+\; \int{d^3p \over (2\pi)^3}
\left( p^2 + m g \rho - {m^2 g^2 \rho^2 \over 2 p^2}
         \right)\,.
\label{I0m1}
\end{eqnarray}
The first integral converges and each
term in the second integral gives a power ultraviolet
divergence. The integral of the $p^2$ term in the second integral 
of~(\ref{I0m1}) 
is independent of $\rho$ and can be cancelled by ${\cal E}(0)$. 
The remaining divergences can
be cancelled in~(\ref{E-rho-ren}) by taking
\begin{eqnarray}
      \delta_1\mu & = & {g\over 4}\int{d^Dp\over(2\pi)^D}\,,
   \label{d1mu}
   \\
\delta_1 g & = & {g^2 m \over 2}
        \int{d^Dp\over(2\pi)^D} {1 \over p^2}\,.
\label{d1g}
\end{eqnarray}
For later convenience, we have generalized the integrals to
an arbitrary number of spatial dimensions $D$.
Similarly we can determine the counterterms $\delta_2\mu$
and $\delta_2 g$ by demanding the cancellation of
$\rho$-dependent power ultraviolet divergences in the
second-order quantum corrections in~(\ref{E-rho-ren}):
\begin{eqnarray}
\delta_2\mu & = & {m g^2\over 4}
         \left( \int{d^Dp\over(2\pi)^D} \right)
         \left( \int{d^Dp\over(2\pi)^D} {1 \over p^2}  \right)\,,
\label{d2mu}
\\
\delta_2 g & = & 
{m^2 g^3 \over 8}
\left( \int{d^Dp\over(2\pi)^D} {1 \over p^2} \right)^2\,.
   \label{d2g}
   \end{eqnarray}

In $D=3$ dimensions, all the power ultraviolet divergences 
can be removed by renormalizations of $\mu$, $g$, and ${\cal E}_0$.
However there is still a logarithmic ultraviolet divergence 
coming from the $J$ term in (\ref{E-rho-ren}). That divergence is the only
obstacle to completing our calculation of the energy density
to second order in the quantum corrections.
The predictive power of quantum field theory lies in the fact
that the same renormalizations must remove the ultraviolet 
divergences from all physical quantities. The 
renormalizations that remove the ultraviolet divergences from 
the  ground state energy density must also remove ultraviolet
divergences from the amplitudes for the low-energy scattering
of atoms in the vacuum.
In the next Section, we calculate quantum corrections to the
amplitudes for scattering of atoms in the vacuum. After
identifying the renormalization that removes the
logarithmic divergence from the energy density, we will
complete the calculation of ${\cal E}$ 
in Section~\ref{renormalization}.

\section{Scattering of Atoms in the Vacuum}
\label{scattering}

In 
this Section, we calculate quantum corrections to the 
$T$-matrix elements for $2\to 2$ scattering and for $3\to 3$
scattering of atoms in the vacuum. 
We determine the renormalizations that are 
necessary to remove ultraviolet divergences from these 
$T$-matrix elements.

\subsection{Perturbative framework}

Atoms in the vacuum can be described by the action (\ref{actpsi})
with the chemical
potential $\mu$ set to 0.  A perturbative framework for calculating 
their scattering amplitudes can be obtained by separating the action 
into a free part and an
interaction part as follows:
\begin{eqnarray}
S_{\rm free}[\psi] &=& 
\int dt\,\int d^3x\;
\psi^\dagger 
        \left( i \partial_t + {\mbox{\boldmath
                        $\nabla$}^2\over 2m} \right) \psi  \,,
\label{Sv-free}
\\
S_{\rm int}[\psi] &=& 
\int dt\,\int d^3x\;
\left( - {1 \over 4} g\, (\psi^\dagger \psi)^2
        - {1 \over 36} g_3\, (\psi^\dagger \psi)^3
        - \ldots \right) \,.
\label{Sv-int}
\end{eqnarray}
We can read off the Feynman propagator from $S_{\rm free}$.  Its Fourier
transform is
\begin{equation}
D(\omega, k) = {i \over \omega - k^2/2m+i \epsilon}
\end{equation}
This propagator is represented by a solid line with an arrow as illustrated in
Fig.~\ref{feynpsi}(a).  The 4-particle and 6-particle 
interactions in (\ref{Sv-int}) are
represented by vertices connecting 4 lines and 6 lines, respectively, as
shown in Fig.~\ref{feynpsi}(b) and~\ref{feynpsi}(c).  These vertices have equal number of arrows 
entering and exiting.  This reflects the conservation of the number 
of atoms, which follows from the phase symmetry 
$\psi \to e^{i \alpha} \psi$ of the
action consisting of (\ref{Sv-free}) and (\ref{Sv-int}).

\subsection{\boldmath{$2\to 2$} scattering}
\label{2to2}

Two atoms with momenta ${\bf k}_1$ and ${\bf k}_2$ can scatter 
into states with momenta ${\bf k}'_1$ and ${\bf k}'_2$ that are
allowed by conservation of energy and momentum. The 
probability amplitude for the scattering process is given by
the $T$-matrix element ${\cal T}( {\bf k}_1,{\bf k}_2;
{\bf k}'_1,{\bf k}'_2 )$.

The only terms in the action that contribute to $2\to 2$ scattering are
those that are fourth order in $\psi$. If the only such term is
$(\psi^\dagger\psi)^2$, the scattering is purely S-wave.
In this case, the $T$-matrix element is a function of a single
variable:
\begin{equation}
      {\cal T}( {\bf k}_1,{\bf k}_2; {\bf k}'_1,{\bf k}'_2 ) =
         {\cal T}(q_{12})\,,
   \end{equation}
where $q_{12}=|{\bf k}_1 - {\bf k}_2|$. The center-of-mass energy
$E$ is related to $q_{12}$ by $E = q_{12}^2/4m$.

The $T$-matrix element for $2\to 2$ scattering can be
expanded in the order of quantum corrections:
\begin{equation}
      {\cal T}(q_{12}) = 
      - g + {\cal T}_1(q_{12}) 
      + {\cal T}_2(q_{12}) + \ldots \,.
   \end{equation}
The first term comes from the tree diagram in
Fig.~\ref{scat:22}(a). The first quantum correction comes
from the one-loop diagram in Fig.~\ref{scat:22}(b). Using
contour integration to evaluate the energy integral, we get
\begin{equation}
   \label{t2a}
      {\cal T}_1(q_{12}) = {mg^2\over 2} 
      \int{d^3 p\over (2\pi)^3}
      {1\over p^2 - {\bf p}\cdot({\bf k}_1+{\bf k}_2) 
      + {\bf k}_1\cdot{\bf k}_2-i\epsilon}   \,.
   \end{equation}
The integral has a linear ultraviolet divergence, but the divergence
can be removed by renormalization of the coupling
constant $g$. Including the counterterms from the tree
diagram in Fig.~\ref{scat:22}(a), we get
\begin{equation}
   \label{t2b}
      {\cal T}_1(q_{12}) = {mg^2\over 2} 
      \int{d^3 p\over (2\pi)^3}
      {1\over p^2 - {\bf p}\cdot({\bf k}_1+{\bf k}_2) 
      + {\bf k}_1\cdot{\bf k}_2-i\epsilon} - \delta_1 g  \,.
   \end{equation}
To maintain rotational symmetry, we must shift the
integration variable by 
${\bf p}\to{\bf p}+({\bf k}_1+{\bf k}_2)/2$ before
imposing an ultraviolet cutoff $|{\bf p}|<\Lambda_{UV}$.
The resulting integral can be evaluated analytically and we
obtain
\begin{equation}
{\cal T}_1(q_{12}) \;=\; {mg^2\over 4\pi^2}\Lambda_{UV}
      \;-\; {mg^2\over 16\pi}(-q_{12}^2 - i\epsilon)^{1/2} 
      \;-\; \delta_1 g \,.
\label{t2c}
\end{equation}
The counterterm $\delta_1 g$, whose value was determined in 
(\ref{d1g}), precisely cancels the linear
ultraviolet divergence in (\ref{t2c}).  The final result is
\begin{equation}
{\cal T}_1(q_{12}) \;=\; i {mg^2\over 16\pi} q_{12} \,.
\label{t2d}
\end{equation}

A particularly convenient method for regularizing
ultraviolet divergent integrals is {\em dimensional
regularization\/}. The number of spatial dimensions $D$ is
taken to be a complex variable. The integral is evaluated as
a function of $D$ in a region of the complex $D$-plane where it
converges. This defines an analytic function of $D$ which
can be analytically continued to $D=3$. After shifting the
integration variable in~(\ref{t2a}) by 
${\bf p}\to{\bf p}+({\bf k}_1+{\bf k}_2)/2$ and then
integrating over the angles in $D$ dimensions, we obtain
\begin{equation}
      {\cal T}_1(q_{12}) = 
      {mg^2\over (4\pi)^{D/2}\Gamma\left({D\over 2}\right)}
      \int_0^{\infty} dp\,
      {p^{D-1}\over p^2 - q_{12}^2/4 - i\epsilon}\,.
   \end{equation}
The integral converges for Re$D<2$ and is given by
\begin{equation}
      {\cal T}_1(q_{12}) = 
      {mg^2\over 2}
      {\Gamma(1-D/2)\over(4\pi)^{D/2}}
      \left({-q_{12}^2\over 4} - i\epsilon\right)^{(D-2)/2} \,.
   \end{equation}
Analytically continuing to $D=3$, we recover the
result~(\ref{t2d}). 

One of the great advantages of dimensional regularization is that
integrals that contain no scale are set identically equal to 0:
\begin{equation}
      \int{d^D p\over(2\pi)^D} \; p^\alpha \;=\; 0
\label{intp0}\, .
\end{equation}
This formula can be derived by first integrating over angles 
in $D$ dimensions, and then separating the integral over $p$ 
into two pieces corresponding to $p < p^*$ and
$p>p^*$.  The integral over $p < p^*$ can be evaluated for $D$ large enough
that it is convergent in the infrared.  
The integral over $p > p^*$ can be evaluated for
$D$ small enough that it is convergent in the ultraviolet.  Upon analytically
continuing the two integrals to $D = 3$, we find that they cancel exactly.  
Because of the identity (\ref{intp0}), dimensional regularization 
sets pure power ultraviolet divergences to 0.
Thus the counterterms $\delta \mu$ and $\delta g$, which are given by
the integrals in  (\ref{d1mu})-(\ref{d2g}), vanish. 
With dimensional regularization,
the only ultraviolet divergences that require explicit 
renormalization are logarithmic divergences, which appear as
poles in $D-3$.  Nontrivial counterterms are therefore needed 
only to cancel logarithmic ultraviolet divergences.

The quantum corrections to the scattering amplitude 
from higher order diagrams, like the
two-loop diagram in Fig.~\ref{scat:22}(c), form a geometric
series and can be summed up exactly. The $n$th term in
the series is
\begin{equation}
   \label{t2g}
      {\cal T}_n(q_{12}) = 
      -g \left(-i{m g \over 16\pi} q_{12}\right)^n \,.
   \end{equation}
Summing up the geometric series, the complete $2\to 2$
scattering amplitude is
\begin{equation}
   \label{t2h}
      {\cal T}(q_{12}) = 
      {-g\over 1 + i m g q_{12}/(16\pi)}\,.
   \end{equation}
The imaginary part of this $T$-matrix element is precisely
that required by the optical theorem.

\subsection{\boldmath{$3\to 3$} Scattering}

Three atoms with momenta ${\bf k}_1$, ${\bf k}_2$
and ${\bf k}_3$ can scatter into states with momenta
${\bf k}'_1$, ${\bf k}'_2$
and ${\bf k}'_3$ that are allowed by conservation of energy
and momentum. The probability amplitude for $3\to 3$
scattering processes in which all 3 atoms participate is
given by the connected $T$-matrix element, which we denote
by ${\cal T}( {\bf k}_1,{\bf k}_2,{\bf k}_3;
{\bf k}'_1,{\bf k}'_2,{\bf k}'_3 )$. For simplicity, we 
consider only the center of momentum frame, where 
${\bf k}_1+{\bf k}_2+{\bf k}_3=0$ and we use the shorthand
\begin{equation}
      {\cal T}(123\to 1'2'3')\equiv
      {\cal T}({\bf k}_1,{\bf k}_2,{\bf k}_3;
         {\bf k}'_1,{\bf k}'_2,{\bf k}'_3 ) \,.
   \end{equation}

The connected $T$-matrix element for $3\to 3$ scattering can be
separated into the terms that involve a single virtual
particle in the intermediate state and the remainder,
which is called the {\em one-particle-irreducible\/} (1PI)
part of ${\cal T}$:
\begin{eqnarray}
   \label{t3}
      {\cal T}( 123\to 1'2'3' ) &=&
      {\cal T}^{\rm 1PI}( 123\to 1'2'3' )
   \nonumber \\ &&
      + \sum_{(123)} \sum_{(1'2'3')}
         {\cal T}(q_{12})
         {m\over {\bf k}_1\cdot{\bf k}_2
            -({\bf k}_1+{\bf k}_2)\cdot{\bf
            k}'_3+k_3^{'2}-i\epsilon}
         {\cal T}(q_{1'2'}) \,,
   \end{eqnarray}
where $q_{12}=|{\bf k}_1 - {\bf k}_2|$ and 
$q_{1'2'}=|{\bf k}'_1 - {\bf k}'_2|$. The sums are over
cyclic permutations of ${\bf k}_1$, ${\bf k}_2$, and 
${\bf k}_3$ and of ${\bf k}'_1$, ${\bf k}'_2$, and 
${\bf k}'_3$. 
The term in the sum
that is given explicitly corresponds to the $2\to 2$
scattering of particles 1 and 2 to produce particle $3'$ and
a virtual particle. A subsequent $2\to 2$ scattering of the
virtual particle and particle 3 produces particles $1'$ and
$2'$. Examples of diagrams that contribute to the sum are
the tree diagram in Fig.~\ref{scat:33}(a) and the one-loop
diagram in Fig.~\ref{scat:33}(b).

The leading contributions to the 1PI $T$-matrix element for
$3\to 3$ scattering come from one-loop diagrams like the one
in Fig.~\ref{scat:33-1PI}(a). After using contour integration to
integrate over the loop energy, we obtain
\begin{equation}
{\cal T}^{\rm 1PI}_1 ( 123\to 1'2'3' ) \;=\;
- m^2 g^3 \sum_{(123)} \sum_{(1'2'3')} {\cal I}(123 \to 1'2'3') \,,
   \label{t31pia}
   \end{equation}
where 
\begin{equation}
{\cal I}(123 \to 1'2'3') \;=\;
\int{d^3p\over(2\pi)^3}
         {1\over (p^2 + {\bf p}\cdot{\bf k}_3 
            + {\bf k}_1\cdot{\bf k}_2-i\epsilon)
         ( p^2 + {\bf p}\cdot{\bf k}'_3
            + {\bf k}'_1\cdot{\bf k}'_2-i\epsilon)} \,.
   \end{equation}
This integral is ultraviolet convergent.

The next most important quantum corrections to the 1PI
$T$-matrix element come from two-loop diagrams, such as those in
Fig.~\ref{scat:33-1PI}(b), \ref{scat:33-1PI}(c), and \ref{scat:33-1PI}(d), 
and from the insertion of a counterterm $\delta_1 g$ into the one-loop
diagram in Fig.~\ref{scat:33-1PI}(a). After using contour
integration to integrate over the loop energies, we obtain
\begin{eqnarray}
{\cal T}^{\rm 1PI}_2( 123\to 1'2'3' ) &=&
      {\delta_1 g\over g}\,
      {\cal T}^{\rm 1PI}_1( 123\to 1'2'3' )
\nonumber \\ 
&& \hspace{-1in}
+ m^2 g^2 \sum_{(123)} \sum_{(1'2'3')}
         \left[ {\cal T}_1(q_{12}) + {\cal T}_1(q_{1'2'}) \right]
         {\cal I}( 123 \to 1'2'3')
\nonumber \\ 
&& \hspace{-1in}
      \;+\; m^3 g^4 \sum_{(123)} \sum_{(1'2'3')}
      \int{d^3p\over(2\pi)^3}
      \int{d^3q\over(2\pi)^3}
      {1\over p^2+q^2+r^2-2mE-i\epsilon}
\nonumber \\ 
&& \hspace{-0.5in}
\times \; \Bigg\{ 
         {2\over (p^2 + {\bf p}\cdot{\bf k}_3 
            + {\bf k}_1\cdot{\bf k}_2-i\epsilon)
         ( q^2 + {\bf q}\cdot{\bf k}'_3 
            + {\bf k}'_1\cdot{\bf k}'_2-i\epsilon)}
\nonumber \\ 
&& \hspace{-0.5in} \qquad
      +{1\over (p^2 + {\bf p}\cdot{\bf k}_3 
            + {\bf k}_1\cdot{\bf k}_2-i\epsilon)
         ( p^2 + {\bf p}\cdot{\bf k}'_3 
            + {\bf k}'_1\cdot{\bf k}'_2-i\epsilon)}
      \Bigg\} \,,
   \label{t31pib}
\end{eqnarray}
where $r = |{\bf p} + {\bf q}|$ 
and $E=(k_1^2+k_2^2+k_3^2)/2m$ is the total energy. 
The integral over ${\bf q}$ in the last term
of~(\ref{t31pib}) has a linear ultraviolet divergence. 
Using the expression~(\ref{d1g}) for the counterterm $\delta_1 g$,
we can see that the first term on the right side 
of~(\ref{t31pib}) cancels the
linear divergence from the integral over ${\bf q}$ in
the last term.

After cancellation of the linear divergence, there remains
an overall logarithmic divergence 
in the integral over ${\bf p}$ and ${\bf q}$ in (\ref{t31pib}).
This is evident from
scaling ${\bf p}\to t\,{\bf p}$ and ${\bf q}\to t\,{\bf
q}$. As $t\to\infty$, the integrand scales like $1/t^6$
while the integration volume scales like $t^6$. The
divergence is independent of the external momenta and
therefore corresponds to a point interaction between the
three particles. The divergence can be cancelled by the
counterterm $\delta g_3$ associated with 
the $(\psi^\dagger\psi)^3$ term
in the action~(\ref{actpsi}).
However, if we include that counterterm,  we must also for
consistency include the contribution to  ${\cal T}(123\to
1'2'3')$ from the coupling constant $g_3$. Thus we must add
to~(\ref{t31pib}) the contribution from the tree diagram in
Fig.~\ref{feynpsi}(c):
\begin{equation}
\Delta{\cal T}
      \;=\; -(g_3 + \delta g_3)\,.
\label{t3point}
\end{equation}

We choose to use dimensional regularization to regularize
the integral in~(\ref{t31pib}). The logarithmic ultraviolet
divergence then appears as a pole in $D-3$. The pole in
the integral  over ${\bf p}$ and ${\bf q}$
in~(\ref{t31pib}) is identical to that of the
following integral, which is evaluated in the limit $D\to 3$
in Appendix~\ref{K1K2}:
\begin{eqnarray}
&&
\int{d^Dp\over(2\pi)^D} \int{d^Dq\over(2\pi)^D}
\Bigg\{ 
{2 \over (p^2 + q^2 + r^2 + 2 \kappa^2) (p^2 + \kappa^2) (q^2 + \kappa^2)}
\nonumber \\ 
&& \qquad\qquad\qquad\qquad
\;+\; {1 \over (p^2 + q^2 + r^2 + 2 \kappa^2) (p^2 + \kappa^2)^2} 
\;-\; {1 \over 2 (q^2 + \kappa^2) (p^2 + \kappa^2)^2} 
\Bigg\}
\nonumber \\ 
&& \qquad\qquad
\;=\; -{4\pi-3\sqrt{3}\over 192\pi^3}
         \left(  {1\over D-3} - 1.13459 \right)
         \kappa^{2(D-3)} \,.
   \label{polet3}
   \end{eqnarray}
The last term in the integrand of~(\ref{polet3}) cancels the
linear divergence in the integral over ${\bf q}$ of the
previous term,
but does not change the logarithmic ultraviolet divergence,
which gives the pole in $D-3$.
The pole term in~(\ref{t31pib}) is therefore
\begin{equation}
   \label{t3pole}
      \left[
         {\cal T}^{\rm 1PI}_2 (123\to 1'2'3')
      \right]_{\rm pole}
      = -{3(4\pi-3\sqrt{3})\over 64 \pi^3 (D-3)} m^3 g^4 \,.
   \end{equation}
The pole must be cancelled by the
counterterm $\delta g_3$ in~(\ref{t3point}). One of the simplest
renormalization schemes is {\em minimal
subtraction\/}~\cite{t'Hooft-Veltman}. This scheme defines a
{\em running coupling constant\/} $g_3(\kappa)$ that depends
on an arbitrary {\em renormalization scale} $\kappa$. 
The minimal subtraction prescription is to choose the counterterm 
to be a pure pole in
$D-3$ multiplied by a power of $\kappa$:
\begin{equation}
   \label{dg3}
      \delta g_3(\kappa) = 
         -{3(4\pi-3\sqrt{3})\over 64\pi^3(D-3)}
         m^3 g^4 \kappa^{2(D-3)} \,.
   \end{equation}
The exponent of $\kappa$
is chosen so that both sides of~(\ref{dg3}) have
the same engineering dimensions even when $D\neq 3$. Without
such a factor, renormalized quantities would involve
logarithms of dimensionful quantities. The power of $\kappa$
in~(\ref{dg3}) is determined by dimensional analysis. With
$\hbar$ set equal to 1, the terms in the action must be
dimensionless.  Time has dimensions $m L^2$ when $\hbar = 1$,
where $L$ refers to length.  The integration 
measure $\int dt \int d^Dx$ therefore has dimensions $mL^{D+2}$.
Since $\psi^\dagger\psi$ is a number density,
$\psi$ has dimensions $[\psi]=L^{-D/2}$. The dimensions of
the coupling constants are then $[g]=L^{D-2}/m$ and
$[g_3]=L^{2D-2}/m$. The power of $\kappa$ in~(\ref{dg3})
provides the extra factor of $L^{-2D+6}$ required for
dimensional consistency.

Physical quantities cannot depend on the arbitrary
parameter $\kappa$ introduced through the
counterterm~(\ref{dg3}). The coupling constant
$g_3(\kappa)$ must therefore depend on $\kappa$ in such a way that
the combination $g_3+\delta g_3$ is independent of $\kappa$.
This statement can be conveniently expressed in the form of
a {\em renormalization group equation\/}.
Using~(\ref{dg3}),
the condition $(d/d\kappa)(g_3+\delta g_3) = 0$ 
reduces in the limit $D\to 3$ to
\begin{equation}
      \kappa{d\ \over d\kappa} g_3 = 
         {3(4\pi-3\sqrt{3})\over 32\pi^3} m^3 g^4 \,.
\label{rgeg3}
\end{equation}
Since $m$ and $g$ are independent of $\kappa$, the solution
to the equation~(\ref{rgeg3}) is
\begin{equation}
      g_3(\kappa') = g_3(\kappa) +
         {3(4\pi-3\sqrt{3})\over 32\pi^3} m^3
            g^4\ln{\kappa'\over\kappa} \,.
\label{rg-sol}
\end{equation}
This equation tells us that the parameter $g_3$ is a
running coupling constant that varies 
logarithmically 
with the renormalization scale $\kappa$.  The renormalization scale
$\kappa$ can be interpreted as the inverse of the spatial
resolution.  As $\kappa$ increases, the spatial resolution becomes finer
and part of the ``pointlike'' $3 \to 3$ scattering amplitude
represented by the diagram in Fig.~\ref{feynpsi}(c) is resolved into
the successive $2 \to 2$ scatterings represented by the
diagrams in Figs.~\ref{scat:33-1PI}(c) and \ref{scat:33-1PI}(d).  
The contributions from the two individual 
two-loop diagrams have opposite signs and the net effect
is that the coupling constant $g_3$ increases as $\kappa$
increases.

The renormalized expression for ${\cal T}^{\rm 1PI}_2(123\to
1'2'3')$ is given by the limit as $D\to 3$ of the sum
of~(\ref{t31pib}) and~(\ref{t3point}). To take the limit,
we must extract the pole in $D-3$ from the 
integral over ${\bf p}$ and ${\bf q}$ in~(\ref{t31pib}), 
so that it can be cancelled by the counterterm $\delta g_3$.
This can be accomplished by subtracting the integrand
on the left side of~(\ref{polet3}) from the 
last integrand in~(\ref{t31pib}), and then adding to~(\ref{t31pib})
the right side of~(\ref{polet3}) multiplied by $9 m^3 g^4$.
The integral  over ${\bf p}$ and ${\bf q}$ 
is then convergent in $D=3$, 
but depends logarithmically on the scale $\kappa$.

We have eliminated the logarithmic ultraviolet divergence from 
${\cal T}^{\rm 1PI}$
by expressing it in terms of the
renormalized coupling constant $g_3(\kappa)$ defined by
dimensional regularization and minimal subtraction.
The resulting expression for ${\cal T}^{\rm 1PI}$
can serve as a definition of $g_3(\kappa)$
that makes no reference to the regularization  scheme.
Someone who prefers a more physical definition of the
coupling constant can define $g_{3, {\rm phys}}$ to be equal to the value of
$-{\cal T}^{\rm 1PI}(123 \to 1'2'3')$ at their favorite configuration
of the initial and final momenta.
Any such coupling constant can be expressed in the form
\begin{equation}
g_{3,{\rm phys}} \;=\; g_3 (\kappa) + C (\kappa) m^3 g^4.
\label{g3-phys}
\end{equation}
Since its definition makes no reference to the renormalization scale 
$\kappa$, the physical coupling constant satisfies 
$(d/d\kappa) g_{3,{\rm phys}} = 0$.
The renormalization group equation (\ref{rgeg3}) 
then implies that the coefficient
$C(\kappa)$ in (\ref{g3-phys}) is a linear function of  $\ln \kappa$.  
It therefore vanishes for some value $\kappa_{\rm  phys}$ 
and we have
\begin{equation}
g_{3,{\rm phys}} \;=\; g_3(\kappa_{\rm phys}) .
\end{equation}
Thus any physical definition of the coupling constant is equivalent to the
running coupling constant $g_3(\kappa)$ evaluated at a particular value 
of the renormalization scale $\kappa$.  Thus there is little to be gained 
by using a more physical definition of the coupling constant.

The integrals in ${\cal T}^{\rm 1PI}_2 + \Delta{\cal T}$ are functions of the 
renormalization scale $\kappa$ and the initial and final momenta.
For momentum configurations in which the squares of 
the momenta and their inner products are all
comparable in magnitude, the only scales in the integrands are 
$\kappa^2$ and $m E$, where $E$ is the total center-of-mass energy.
Since the integral varies logarithmically with $\kappa$,
it must also depend logarithmically on $mE$.
If the ratio of these two scales is sufficiently large, the
integral is dominated by the logarithm.
However, since the dependence of the logarithm on $\kappa$ is cancelled 
by $g_3(\kappa)$,  there must be a large cancelling contribution from
the $g_3$ term.  The dominant terms in the second order 
quantum correction to the $T$-matrix element are therefore
\begin{equation}
{\cal T}^{\rm 1PI}_2 \;+\; \Delta{\cal T}
\;\approx\; - g_3(\kappa)
\;-\; {3 (4\pi - 3\sqrt{3}) \over 64 \pi^3}
         m^3 g^4 \ln{m E \over \kappa^2} \,.
\label{large-log}
\end{equation}
The large logarithm can be avoided by choosing the 
renormalization scale of the running coupling constant
$g_3(\kappa)$ to be $\kappa = \sqrt{m E}$.  Thus
the most appropriate choice for the renormalization scale 
$\kappa$ in the $3 \to 3$ scattering amplitude is the magnitude
of the typical momenta of the scattering particles.
This choice will avoid a large cancellation between the two terms in
(\ref{large-log}).

\section{Renormalized Energy Density}
\label{renormalization}

In this Section, we complete the calculation of the energy
density in Section~\ref{EdensityinD} by using the
renormalization of $g_3$ to remove the logarithmic
ultraviolet divergences from the second-order quantum
correction.  Our final result is expressed in terms of
parameters $g$ and $g_3(\kappa)$ that can be defined purely
in terms of the scattering of atoms.  We then discuss the
case of alkali atoms, and argue that in this case the
dependence on $g_3$ can be eliminated in favor of a
logarithmic dependence on the length scale set by the van
der Waals potential.

\subsection{First-order quantum correction}

To simplify the calculation of the quantum corrections, 
we use dimensional regularization to regularize
ultraviolet divergences and minimal subtraction to carry out the 
renormalization of the parameter $g_3$. 
One of the great advantages of dimensional regularization is that it 
sets power ultraviolet divergence to 0. 
The only ultraviolet divergences that require explicit 
renormalization are logarithmic divergences, which appear as
poles in $D-3$.  Since the ultraviolet divergences removed by the 
renormalization of $\mu$ and $g$ are power divergences, 
the explicit renormalization of these parameters is unnecessary
with dimensional regularization.  This can be seen from the expressions 
(\ref{d1mu})--(\ref{d2g}) for the counterterms, which all vanish by
the identity (\ref{intp0}). 
Thus the expression~(\ref{E-rho-ren}) for the energy density
after renormalization of $\mu$ and $g$ collapses to~(\ref{Erho}). 
The integration constant ${\cal E}_0$,
which is used to set ${\cal E}(0) = 0$, is also zero in dimensional
regularization.  This follows from the fact that the
only scale in the dimensionally regularized integrals
$I_{m,n}$ and $J$ is $2 m g \rho$.  These integrals therefore vanish when 
$\rho=0$, since there is no momentum scale.

From~(\ref{Erho}),
the first order quantum correction to the energy density is
\begin{equation}
      {\cal E}_1(\rho) = {1\over 4m}I_{0,-1}(2mg\rho)
\label{E1-rhoA}\,.
\end{equation}
Using the expression for the dimensionally regularized 
integral given in~(\ref{ia}), this becomes
\begin{equation}
      {\cal E}_1(\rho) = {1\over 60\pi^2}
         {(2mg\rho)^{5/2}\over m} \,.
\label{E1-rhoB}
\end{equation}
Using (\ref{g-a}) to express $g$ in terms of the scattering length $a$,
we recover the classic result for the first order
quantum correction given in (\ref{E-Yang}).

\subsection{Second-order quantum correction}

Dimensional regularization eliminates the power ultraviolet
divergences from  the second-order quantum 
correction in~(\ref{Erho}), but the expression still contains 
a logarithmic ultraviolet divergence.
The renormalization of this divergence requires including
the effects of the coupling constant $g_3$ and its 
counterterm at tree level. The term that must be added to
the free energy density can be read off from the integrand
in the action~(\ref{actpsi}):
\begin{equation}
\Delta{\cal F}(\mu) \;=\; 
         {1\over 36}(g_3+\delta g_3) {\overline{v}_0}^6 \,.
\end{equation}
Following the effect of the correction through to the 
energy density, we find
\begin{equation}
\Delta{\cal E}(\rho) \;=\;
        {1\over 36}(g_3+\delta g_3)\rho^3 \,.
\label{dE-rho}
\end{equation}

With dimensional regularization, the complete second-order
quantum correction is the sum of~(\ref{dE-rho}) and the 
correction in~(\ref{Erho}):
\begin{equation}
{\cal E}_2(\rho) \;=\;
         {1\over 36}
         \left[ 
            g_3(\kappa) + \delta g_3(\kappa)
         \right]
         \rho^3
         + {mg^2\rho\over 16}J
         + {g\over 32}
         \left(
            I_{-1,-1}^2 - 2 I_{-1,-1}I_{1,1} - I_{1,1}^2
         \right)\,.
\label{E2-rhoA} 
\end{equation}
The values of the integrals $J$, $I_{-1,-1}$, and $I_{1,1}$
in dimensional regularization are given in~(\ref{numsig}),
(\ref{ib}), and~(\ref{ic}) with $\Lambda^2=2mg\rho$. 
The pole in $D-3$ in $J$ is 
cancelled by the pole in the counterterm $\delta g_3(\kappa)$
given in~(\ref{dg3}), but the cancellation leaves a logarithm
of  $2mg\rho/\kappa^2$ in the limit $D\to 3$. Combining all
the terms, our expression for ${\cal E}_2$ is
\begin{equation}
{\cal E}_2(\rho) \;=\; {1\over 36} g_3(\kappa)\rho^3
      + {4\pi-3\sqrt{3}\over 768 \pi^3}
         \left( \ln{2 m g \rho \over \kappa^2} + 0.80 \right)
         m^3 g^4 \rho^3  \,,
\label{E2-rhoB}
\end{equation} 
where $g_3(\kappa)$ is the running coupling constant defined by 
the minimal subtraction renormalization prescription. 
The expression (\ref{E2-rhoB}) is independent of the arbitrary 
renormalization scale $\kappa$.  
The renormalization group equation~(\ref{rgeg3}) implies that
the explicit logarithmic dependence of~(\ref{E2-rhoB}) is cancelled
by the $\kappa$-dependence of $g_3(\kappa)$.
If $\kappa^2$ differs by orders of magnitude
from $2 m g \rho$, 
there is a large cancellation between the logarithm in (\ref{E2-rhoB})
and the term containing $g_3(\kappa)$. 
Such a large cancellation can be
avoided by choosing the renormalization scale to be 
$\kappa = \sqrt{2 m g \rho}$.   
This is the momentum scale at which the dispersion relation for the 
Bogoliubov mode changes from linear to quadratic.
Our final result for the energy density to second order in the
quantum corrections is obtained by adding the
corrections~(\ref{E1-rhoB}) and~(\ref{E2-rhoB}) to
the mean-field contribution:
\begin{eqnarray}
{\cal E}(\rho) & = & {1\over 4} g \rho^2
         \;+\; {1\over 60\pi^2} {(2mg)^{5/2}\over m} \rho^{5/2}
\nonumber \\ &&
\;+\; {1\over 36}
\left[ g_3(\kappa)
         + {3(4\pi-3\sqrt{3})\over 64 \pi^3}
            \left(\ln{2 m g \rho \over \kappa^2} + 0.80 \right)
            m^3 g^4
      \right]
      \rho^3  \,,
\label{final}
\end{eqnarray}

It would be very difficult to measure the coupling constant $g_3(\kappa)$
experimentally by studying the 3-body scattering of atoms.  
It would also be difficult to calculate this parameter theoretically 
from a microscopic description of the interaction between atoms.  
Thus $g_3(\kappa)$ must be treated as a phenomenological 
parameter.  The predictive power of the result (\ref{final}) 
for the energy density resides in the fact that the same coupling constant
$g_3(\kappa)$ appears in the second order quantum corrections 
to other physical quantities, such as the dispersion relation 
for the Bogoliubov modes.  All of the low-energy observables 
of the Bose gas can be calculated to second order in the
quantum corrections in terms of two phenomenological parameters:  
the S-wave scattering length $a$ and the coupling constant 
$g_3(\kappa)$ associated with $3 \to 3$ scattering.

\subsection{Alkali atoms}

Our final result (\ref{final}) for the energy density is of
limited utility because it depends on the phenomenological
parameter $g_3 (\kappa)$.  It may be possible to neglect the
dependence on $g_3$ if the second quantum correction is
dominated by the logarithm. We argue that this is indeed the
case for typical alkali atoms, and that the dependence on
$g_3$ can be eliminated in favor of a logarithmic dependence
on the length scale set by the van der Waals interaction.

The interaction between two atoms at low energy can be
described by a potential $V(r)$ that has a repulsive core at
very short distances, an attractive region at short
distances comparable to the size of the atom, and a
long-range behavior given by the van der Waals potential:
\begin{equation}
  V(R)\rightarrow -{\alpha\over R^6} \;. 
  \label{vdWpot}
\end{equation}
The scattering length $a$ is extremely sensitive to the
short-distance behavior of the potential. Small variations
in the depth or range of the potential can easily cause $a$
to vary from $-\infty$ to $+\infty$. However, given a random
distribution in one of the short-distance parameters of the
potential, the distribution of $a$ is concentrated in the
region where $|a|$ is less than or comparable to the van der
Waals length defined by~\cite{G-F}
\begin{equation}
    \ell_V=\left({m\alpha\over 9.58}\right)^{1/4} 
  \;.
  \label{g-f}
\end{equation}
For most values of the short-distance parameters, $a$ is
comparable in magnitude to $\ell_V$.  However, $a$ varies
dramatically with the short-distance parameters near the
critical values at which a new 2-body bound state appears.
As a parameter passes through its critical value, $a$
approaches $\pm \infty$, changes discontinuously to $\mp
\infty$, and then decreases in magnitude.  The magnitude of
$a$ will be orders of magnitude larger than $\ell_V$ only if
the short-distance parameter is in a narrow range around its
critical value.  We summarize this situation by saying that
the {\it natural magnitude} of $a$ is $\ell_V$, and that
much larger values require {\it fine-tuning} of the
potential.

One can use the concept of natural magnitudes
to estimate the magnitude of the coupling constants in the lagrangian
for an effective field theory.   A coupling constant 
with dimensions $L^n/m$, where  $L$ refers
to length, will by dimensional analysis have the form 
$f \ell_V^n/m$, where $f$ is dimensionless.  
The assumption of naturalness is that the coefficient 
$f$ is of order 1 except when a short-distance parameter of
the potential is tuned to within a 
narrow range of a critical value.  Since the
coupling constants $g$ and $g_3$ in the action (\ref{actpsi}) 
have dimensions $L/m$ and $L^4/m$, respectively, the natural 
estimates for their magnitudes are $|g| \sim \ell_V/m$ and 
$|g_3| \sim \ell_V^4/m$, respectively.

One can improve on these naive estimates by taking into
account geometrical factors of $4 \pi$.  For example, using
the relation~(\ref{g-a}) between the coupling constant and the
scattering length $a$, we obtain the
estimate
\begin{equation}
| g |_{\rm natural} \;\sim\; {8 \pi \ell_V \over m}  \,.
\label{g-nat}
\end{equation}
We argue that the magnitude of $g$ should be comparable to this natural
estimate unless the 2-body potential is tuned so that 
there is a bound state near threshold.

We next consider the coupling constant $g_3(\kappa)$ associated with 
$3 \to 3$ scattering.  This is a running coupling constant that 
depends on an arbitrary renormalization scale $\kappa$.  
An estimate of the magnitude of this coupling constant 
must include a specification of the scale $\kappa$ at which 
the estimate applies.  Our estimates of natural values, 
which involve dimensional analysis, are based on the assumption that 
$\ell_V$ is the only important length scale.  Thus the estimate must 
apply for momentum scales $\kappa$ that correspond to the
length scale $\ell_V$.  We will therefore assume that the estimate 
of the natural value applies to $g_3(\kappa)$ for $\kappa$
comparable to $2 \pi/ \ell_V$.  Making a guess for the
appropriate values of $4\pi$, our estimate for the natural
value is  
\begin{equation}
\left| g_3(\kappa) \right|_{\rm natural}
        \;\;\sim\;\; 
{(4 \pi \ell_V)^4 \over m} 
\qquad {\rm for} \; \kappa \sim {2 \pi \over \ell_V}.
\label{g3-nat}
\end{equation}
We will verify that this guess passes a simple consistency
check. If both $g$ and $g_3$ have natural values, then we
would expect their values not to change dramatically under
changes of the renormalization scale by a factor of 2 or 3.
Using the solution (\ref{rg-sol}) to the renormalization
group equation for $g_3(\kappa)$, we see that the change in
$g_3$ from the evolution of $\kappa$
by a factor of $e$ is
\begin{equation}
\Delta g_3
        \;\sim\; 
{3 (4 \pi - 3 \sqrt 3) \over 32 \pi^3} m^3 g^4
\;.
\label{g3-rg}
\end{equation}
If $g$ has the natural magnitude given in (\ref{g-nat}), the
estimate (\ref{g3-nat}) is approximately equal to
(\ref{g3-rg}), which indicates that our guess of the factors
of $4 \pi$ in (\ref{g3-nat}) is at least reasonable.

We now consider an atom for which $g_3$ has the natural
magnitude given by~(\ref{g3-nat}). It could be much larger
if the potentials describing 2-body and 3-body interactions
are tuned so that there is a 3-body bound state near
threshold. We cannot exclude such a possibility, but if we
pick an alkali atom at random it is unlikely. We proceed to
consider the three cases where $g$ is much smaller than,
comparable to, and much greater than the natural estimate
given in~(\ref{g-nat}). If $g$ is much smaller than
$|g|_{\rm natural}$, then renormalization has little effect
on the value of $g_3$. The $m^3 g^4$ term in the second
order correction to the energy density in~(\ref{final}) is
negligible compared to the $g_3$ term. In this case, we
cannot calculate the second order correction without knowing
the value of $g_3$. Next we consider the case where $g$ is
comparable to $|g|_{\rm natural}$. The $g_3$ term
in~(\ref{final}) is then comparable in magnitude to the
constant term multiplying $m^3 g^4$. These two terms can be
neglected only if the coherence length $(2mg\rho)^{-1/2}$ is
orders of magnitude larger than $\ell_V/2\pi$. In this case,
the logarithmic term dominates and we obtain an estimate of
the second order quantum correction that is independent of
the unknown constant $g_3$:
\begin{equation}
{\cal E} (\rho) \;\approx\; {1 \over 4} g \rho^2 
\;+\; {1 \over 60 \pi^2} {(2 m g)^{5/2} \over m} \rho^{5/2} 
\;+\; {4 \pi - 3 \sqrt 3 \over 768 \pi^3} 
        \left( \ln {mg\rho \ell_V^2 \over 2 \pi^2} \right) 
        m^3 g^4 \rho^3 \,.
\label{final-lsl}
\end{equation}
Finally, we consider the case where $g$ is much larger than
$|g|_{\rm natural}$. In this case, $g_3$ will quickly evolve
under renormalization to a value comparable in magnitude
to~(\ref{g3-rg}). Again we find that the $g_3$ term
in~(\ref{final}) is comparable in importance to the constant
under the logarithm. If the coherence length is orders of
magnitude larger than the van der Waals length, these terms
are negligible compared to the logarithmic term and the
expression~(\ref{final}) for the free energy density reduces
again to~(\ref{final-lsl}). In summary, the approximation in
(\ref{final-lsl}) is valid provided that $g_3$ is not
unnaturally large compared to the estimate~(\ref{g3-nat}),
that $a$ is not unnaturally small compared to $\ell_V$, and
also that
\begin{equation}
  \left| \ln ({\rho a\, \ell_V^2})\right| \; \gg \; 1 \,.
  \label{largelog}
\end{equation}
Under these conditions, we can eliminate the dependence on
$g_3$ in favor of a logarithmic dependence on $\ell_V$.
Using (\ref{g-a}) to express (\ref{final-lsl}) in terms of
the scattering length $a$, we obtain the expression for the
energy density given in (\ref{E-lsl}).

\section {Conclusion}

We have calculated the second order quantum correction to
the energy density of a homogeneous Bose gas.  This is the
first correction that depends on an atomic physics parameter
other than the S-wave scattering length $a$.  We identify
that parameter as a coupling constant $g_3$ that specifies
the point-like contribution to the $3 \to 3$ scattering of
atoms in the vacuum.  The result for the energy density in
terms of $g$ and the running coupling constant $g_3(\kappa)$
is given in (\ref{final}).  In the case of alkali atoms, we
argued that the dependence on $g_3$ can be
eliminated in terms of a logarithmic dependence on the
length scale $\ell_V$ set by the van der Waals
interaction.  The resulting expression for the energy density
is given in (\ref{E-lsl}).

Thus far the only alkali atoms for which Bose-Einstein
condensation has been successfully carried out are
$^{87}$Rb, $^{23}$Na, and $^7$Li.  Our calculations apply
only to atoms with positive scattering length, such as
$^{87}$Rb and $^{23}$Na.  The scattering lengths for these
atoms are $60 \pm 15$~\AA \ for $^{87}$Rb~\cite{Gardner} and
$29 \pm 3$~\AA \ for $^{23}$Na~\cite{Tiesinga}. The
parameter $\alpha$ in the van der Waals
potential~(\ref{vdWpot}) is roughly 7~keV\AA$^6$ for Rb and
1.8~keV\AA$^6$ for Na. The van der Waals length
$\ell_V$ defined in~(\ref{g-f}) is therefore about 60~\AA \ and
30~\AA, respectively. In both cases, the scattering length
is comparable in magnitude to $\ell_V$, so that one of the
conditions for~(\ref{final-lsl}) is satisfied. The
condition~(\ref{largelog}) depends on the density $\rho$ and
will be satisfied if $\rho$ is orders of magnitude larger
than $1/(a\,\ell_V^2)$.

The magnitude of the quantum corrections increases with the
number density.  To see roughly how important these
corrections are in existing magnetic traps, we evaluate them
for the typical densities at the centers of the traps in the
earliest experiments~\cite{BEC-Rb,BEC-Na}.  For $^{87}$Rb
atoms with number density $\rho = 3 \times 10^{12}/{\rm
  cm}^3$, the correction factor in (\ref {E-lsl}) is $1 +
0.004 - 0.0002$.  For $^{23}$Na with $\rho = 3 \times
10^{14}/{\rm cm}^3$, the quantum correction factor in the
energy density is $1 + 0.01 - 0.002$.  In both cases, the
second order quantum correction is an order of magnitude
larger than one would have guessed by squaring the
first-order quantum correction.  The second order correction
is relatively large because the logarithm in (\ref {E-lsl})
is large, having the value $-14$ for $^{87}$Rb and $-12$ for
$^{23}$Na.  Thus the condition~(\ref{largelog}) for the
validity of (\ref {E-lsl}) is indeed satisfied.

The estimates given above suggest that the number densities in the 
Bose-Einstein condensates that have been
produced thus far are not sufficiently high for the effects of
quantum fluctuations on the energy density to be measurable.
Since the first quantum correction scales like $\sqrt{\rho}$,
the quantum corrections can be made larger by increasing the number of 
atoms in the trap.  Unfortunately, the peak density $\rho$ scales like 
$N^{2/5}$ \cite{Baym-Pethick}, so $N$ must be increased by orders of 
magnitude before the effects of quantum fluctuations 
on the energy density will be measurable.  There are however other 
observables that may be more sensitive to the effects of 
quantum fluctuations.  These effects may also be more important
at temperatures near the phase transition for Bose-Einstein
condensation.  We hope that our explicit calculation of 
second-order quantum corrections for the energy density of a homogeneous
Bose gas will stimulate further work on quantifying the effects of 
quantum fluctuations on experimentally measurable observables.

\bigskip

\section*{Acknowledgments}

This work was supported in part by the U.~S. Department of Energy,
Division of High Energy Physics, under Grant DE-FG02-91-ER40690.

\appendix
\bigskip\renewcommand{\theequation}{\thesection\arabic{equation}}

\section{Loop Integrals for the Energy Density}
\setcounter{equation}{0}\label{ijs}

In this appendix, we evaluate the integrals that are needed to calculate the 
second order quantum corrections to the energy density.
We also list the expressions for each of the two-loop diagrams.

\subsection{Energy integrals}
\label{energyint}

The energy integrals can be evaluated by contour integration.
The energy integral for the one-loop subdiagrams in 
Figs.~\ref{twoloop}(a)--\ref{twoloop}(c) are
\begin{equation}
\int{d\omega\over 2\pi}
      {1\over \omega^2-\varepsilon^2(p)+i\epsilon}
      \;=\; - {i\over 2\varepsilon(p)} \, ,
\end{equation}
where $\varepsilon(p)$ is the Bogoliubov dispersion relation 
given in (\ref{Bogol}).
The energy integrals for the two-loop diagrams in
Figs.~\ref{twoloop}(d)--\ref{twoloop}(g) are
\begin{eqnarray}
&&
\int{d\omega_1\over 2\pi} \int{d\omega_2\over 2\pi} \;
{1 \over [\omega_1^2 - \varepsilon^2(p) + i\epsilon]
        [\omega_2^2 - \varepsilon^2(q) + i\epsilon]
        [(\omega_1 + \omega_2)^2 - \varepsilon^2(r) + i\epsilon]}
\nonumber \\ 
&& \qquad\qquad\qquad
\;=\; {1\over 4 \varepsilon(p)\varepsilon(q)\varepsilon(r)
         \left[ \varepsilon(p)+\varepsilon(q)+\varepsilon(r) \right]}\,,
\\ 
&&
\int{d\omega_1\over 2\pi} \int{d\omega_2\over 2\pi} \;
{\omega_1  \omega_2 \over 
        [\omega_1^2 - \varepsilon^2(p) + i\epsilon]
        [\omega_2^2 - \varepsilon^2(q) + i\epsilon]
        [(\omega_1 + \omega_2)^2 - \varepsilon^2(r) + i\epsilon]}
\nonumber \\ 
&& \qquad\qquad\qquad
\;=\; {1\over 4 \varepsilon(r)
         \left[ \varepsilon(p)+\varepsilon(q)+\varepsilon(r) \right]}\,.
   \end{eqnarray}

\subsection{One-loop momentum integrals}
\label{oneloopint}

The one-loop and two-loop corrections to the ground state energy
density involve momentum integrals of the form
\begin{equation}\label{imn}
I_{m,n} \;=\; 
        \int_{\bf p} {p^{2m-n}\over (p^2+\Lambda^2)^{n/2}} \, ,
\end{equation}
where we have introduced the following notation for the integration 
over a momentum in $D$ spatial dimensions: 
\begin{equation}
\int_{\bf p} \;\equiv\;  \int {d^D p\over (2\pi)^D} \, .
\end{equation}
The integrals (\ref{imn}) satisfy the identities
\begin{eqnarray}
{d\ \ \over d\Lambda^2} I_{m,n} & = & - {n \over 2} \; I_{m+1,n+2} \,, 
\label{prop1}
\\
\Lambda^2 I_{m,n} &=& I_{m-1,n-2} - I_{m+1,n} \,.  
\label{prop3}
\end{eqnarray}
After integrating over angles, the integral is
\begin{equation}
I_{m,n} \;=\; {1\over(4\pi)^{D/2} \Gamma({D \over 2})}
         \int_0^{\infty} dp^2 \,
            {(p^2)^{m+(D-n-2)/2}\over(p^2+\Lambda^2)^{n/2}}\,.
\label{Imn-p}
\end{equation}
If the integral is convergent, we can use integration by parts
to derive the identity 
\begin{equation}
(D + 2m - n) I_{m,n} \;=\;  n  I_{m+2,n+2} \,.  
\label{prop2}
\end{equation}

The integral (\ref{Imn-p}) is ultraviolet divergent if 
$D \ge 2(n-m)$ and infrared divergent if $D \le n-2m$.
If the integral converges, its value is
\begin{equation}
I_{m,n} \;=\;
        {1\over(4\pi)^{D/2}} 
        {\Gamma({2n-2m-D \over 2}) \Gamma({D+2m-n \over 2})
               \over \Gamma({n \over 2}) \Gamma({D \over 2})} \, 
        \Lambda^{D+2(m-n)}\,.
\label{Imn-D}
\end{equation}
If the integral (\ref{Imn-p}) is ultraviolet or infrared divergent,
it can be regularized using dimensional regularization.
The regularized integral is obtained by analytically continuing
the expression (\ref{Imn-D}) to $D = 3$.  The result is
\begin{equation} 
I_{m,n} \;=\;
        {1 \over 4 \pi^2} 
        { \Gamma({2n-2m-3 \over 2}) \Gamma({3+2m-n \over 2})
               \over \Gamma({n \over 2}) } \, 
        \Lambda^{3+2(m-n)}\,.
\label{Imn-3}
\end{equation}
The one-loop integrals  that appear in the ground
state energy density are $I_{0,-1}$, $I_{-1,-1}$, and
$I_{1,1}$. In 3 dimensions, these integrals have power ultraviolet
divergences. With dimensional regularization, they are given 
by the finite expressions
\begin{eqnarray}
  I_{0,-1} & = & {1\over 15 \pi^2} \Lambda^5 \,,
\label{ia} 
\\ 
  I_{-1,-1} & = & - {1\over 6 \pi^2} \Lambda^3 \,,
\label{ib} 
\\
  I_{1,1} & = & {1\over 3 \pi^2} \Lambda^3 \,.
\label{ic} 
\end{eqnarray}

\subsection{Two-loop momentum integrals}
\label{twoloopint}

The two-loop correction to the ground state energy
density~(\ref{Erho}) involves a linear combination of the integrals 
\begin{equation} \label{jlmn}
J_{l,m,n} \;=\; \int_{\bf p} \int_{\bf q} 
{ \left( p/\sqrt{p^2+\Lambda^2} \right)^l 
        \left( q/\sqrt{q^2+\Lambda^2} \right)^m
        \left( r/\sqrt{r^2+\Lambda^2} \right)^n
\over p \sqrt{p^2+\Lambda^2} + q \sqrt{q^2+\Lambda^2}
        + r \sqrt{r^2+\Lambda^2}} \, ,
\end{equation}
where $r = |{\bf p} + {\bf q}|$.
In $D=3$, these integrals have quartic and quadratic ultraviolet 
divergences that cancel in the combination of integrals
$J$ given in (\ref{J-def}).
The expression for $J$ can be written
\begin{eqnarray}
J &=& 
\int_{\bf p} \int_{\bf q} 
{1 \over p \sqrt{p^2 + \Lambda^2} + q \sqrt{q^2 + \Lambda^2}
        + r \sqrt{r^2 + \Lambda^2}}
\left[  {6 p \over \sqrt{p^2 + \Lambda^2}}
        \;-\;  {2 \sqrt{p^2 + \Lambda^2} \over p}
\right.
\nonumber \\ 
&& \hspace{1in}
\left.
\;-\; {3 p q r \over \sqrt{p^2 + \Lambda^2} \sqrt{q^2 + \Lambda^2}
                \sqrt{r^2 + \Lambda^2}}
\;-\; {p \sqrt{q^2 + \Lambda^2} \sqrt{r^2 + \Lambda^2}
                \over q r \sqrt{p^2 + \Lambda^2}} \right] \,.
\label{J-int}
\end{eqnarray}
This integral still has linear and logarithmic ultraviolet divergences.
By subtracting and adding appropriate terms in the integrand of $J$,
we can isolate the linear and logarithmic divergences 
into separate terms:
\begin{equation} 
J \;=\;  J_{\rm lin} \;+\; J_{\rm log} \;+\;  J_{\rm num}\, .
\label{J-sep}
\end{equation}
The term containing the linear ultraviolet
divergence is
\begin{equation}   
J_{\rm lin} \;=\; 2 \int_{\bf p}
\left[ 2 - {p \over \sqrt{p^2+\Lambda^2}} 
         - {\sqrt{p^2+\Lambda^2} \over p} \right]
\int_{\bf q} {1\over q^2} \,.
\label{J-lin}
\end{equation}
The  term in (\ref{J-sep}) containing the logarithmic
ultraviolet divergence is
\begin{eqnarray}
J_{\rm log} &=&
- \Lambda^4 \int_{\bf p} \int_{\bf q} \Bigg\{
    {2 \over (p^2+\Lambda^2)(q^2+\Lambda^2)(p^2+q^2+k^2+2\Lambda^2)} 
\nonumber \\ 
&& \qquad\qquad
+ \left[ {1 \over p^2+q^2+k^2+2\Lambda^2}
            - {1 \over 2 (q^2+\Lambda^2)} \right]
            {1 \over (p^2+\Lambda^2)^2}
      \Bigg\} \,.
\label{J-log}
\end{eqnarray}
The integral $J_{\rm num}$ obtained by subtracting (\ref{J-lin})
and (\ref{J-log}) from (\ref{J-int}) is convergent in $D=3$ dimensions
and can be evaluated numerically.
It is convenient to symmetrize the integrand over the six 
permutations of $p$, $q$, and $r$ in order to avoid cancellations 
between different regions of momentum space.  
The resulting expression is
\begin{eqnarray}
J_{\rm num} &=& 
 \int_{\bf p} \int_{\bf q} {1 \over 6} \sum_{(pqr)}
\Bigg\{ 
{1 \over p \sqrt{p^2 + \Lambda^2} + q \sqrt{q^2 + \Lambda^2}
        + r \sqrt{r^2 + \Lambda^2}}
\left[ {6 p \over \sqrt{p^2 + \Lambda^2}}
        - {2 \sqrt{p^2 + \Lambda^2} \over p}
\right.
\nonumber \\ 
&& \hspace{1.5in}
\left.
        - {3 p q r \over \sqrt{p^2 + \Lambda^2} \sqrt{q^2 + \Lambda^2}
                \sqrt{r^2 + \Lambda^2}}
        - {p \sqrt{q^2 + \Lambda^2} \sqrt{r^2 + \Lambda^2}
                \over q r \sqrt{p^2 + \Lambda^2}} \right]
\nonumber \\ 
&& \hspace{0.5in}
\;+\; {2\over q^2}
        \left[ 2   - {p \over \sqrt{p^2 + \Lambda^2}}
                   - {\sqrt{p^2 + \Lambda^2} \over p} \right]
\nonumber \\ 
&& \hspace{0.5in}
\;+\; {2 \Lambda^4 \over (p^2 + q^2 + r^2 + 2\Lambda^2) (p^2 + \Lambda^2) 
                (q^2 + \Lambda^2)}
\nonumber \\ 
&&  \hspace{0.5in}
\;+\; \left[ {1 \over p^2 + q^2 + r^2 + 2 \Lambda^2}
                - {1\over 2 (q^2+\Lambda^2)} \right]
         {\Lambda^4 \over (p^2+\Lambda^2)^2}
\Bigg\} \,.
\label{J-num}
\end{eqnarray}
Since $\Lambda$ is the only scale in the integrand, dimensional 
analysis implies that the integral is proportional to 
$\Lambda^4$. 
Evaluating the coefficient of $\Lambda^4$ numerically, we
obtain
\begin{equation}
J_{\rm num} \;=\; 2.10 \times 10^{-3}  \; \Lambda^4 \,.
\label{J-1}
\end{equation}
Because of the severe cancellations between the various terms in the
integral,  we were only able to calculate it
to 3  significant figures.

We evaluate the ultraviolet divergent integrals $J_{\rm lin}$
in (\ref{J-lin}) and $J_{\rm log}$ in (\ref{J-log}) using 
dimensional regularization.  The integral 
over ${\bf q}$ in (\ref{J-lin}) vanishes
since there is no scale in the integrand, and therefore
$J_{\rm lin} = 0$.
The integral  (\ref{J-log}) is evaluated in Appendix \ref{K1K2}
in the limit $D \to 3$, and the result is
\begin{equation}
J_{\rm log} \;=\; {4\pi - 3\sqrt{3} \over 192 \pi^3} 
\left( {1 \over D-3} - 1.13459 \right) \Lambda^{4+2(D-3)} \,.
\label{J-3}
\end{equation}
Adding~(\ref{J-1}) and~(\ref{J-3}), we obtain
the complete result for $J$ using dimensional regularization:
\begin{eqnarray}
J \;=\; {4\pi - 3\sqrt{3} \over 192 \pi^3} 
\left( {1\over D-3} + 0.57 \right) \Lambda^{4+2(D-3)} \,.
\label{numsig}
\end{eqnarray}

\subsection{Evaluation of $J_{\rm log}$}
\label{K1K2}

The term (\ref{J-log}), which contains the logarithmic 
ultraviolet divergence in the integral $J$, can be written
\begin{equation}
     J_{\rm log} \;=\; - \Lambda^4 \, ( 2 K_1 + K_2 ) \,,
\end{equation}
where $K_1$ and $K_2$ are the following integrals:
\begin{eqnarray}
K_1 &=& \int_{\bf p} \int_{\bf q} 
{1 \over (p^2+q^2+r^2+2\Lambda^2)(p^2+\Lambda^2)(q^2+\Lambda^2) }\,,
\label{K1-def}
\\
K_2 &=& \int_{\bf p} \int_{\bf q}
\left[ {1 \over p^2 + q^2 + r^2 + 2\Lambda^2} 
        - {1 \over 2 (q^2 + \Lambda^2)} \right]
        {1 \over (p^2+\Lambda^2)^2}\,,
\label{K2-def}
\end{eqnarray}
where $r=|{\bf p}+{\bf q}|$. 

We first consider the integral $K_1$.
Setting $r^2 = p^2 + q^2 + 2 {\bf p} \cdot {\bf q}$
and then introducing Feynman parameters, the integral (\ref{K1-def})
becomes
\begin{equation}
K_1 \;=\; \int_0^1 dx \int_0^{1-x} dy 
\int_{\bf p} \int_{\bf q} { 1 \over [ (1-y)p^2 + (1-x)q^2 
        + z {\bf p} \cdot {\bf q} + \Lambda^2 ]^3 } \,,
\end{equation}
where $z=1-x-y$. Dimensional regularization allows us to shift 
and rescale the momentum variables.  
We can eliminate the dot product from the denominator
by making the shift ${\bf p} \to {\bf p} - {z \over 2(1-y)}{\bf q}$.
After rescaling ${\bf p}$ by $(1-y)^{-1/2}$ and 
${\bf q}$ by $\left( {(1-x)(1-y)-z^2/4 \over 1-y} \right)^{-1/2}$,
the integral factors into a Feynman parameter integral and an 
integral over the momenta:
\begin{equation}
K_1 \;=\; \int_0^1 dx \int_0^{1-x} dy 
[ (1-x)(1-y) - z^2/4 ]^{-D/2}
\int_{\bf p} \int_{\bf q} {1 \over (p^2 + q^2 + \Lambda^2)^3 } \,.
\end{equation}
The integral over the momenta can be evaluated analytically:
\begin{equation}
\int_{\bf p} \int_{\bf q} {1 \over (p^2 + q^2 + \Lambda^2)^3 }
 \;=\; {\Gamma(3-D) \over 2 (4 \pi)^D} \Lambda^{2(D-3)}\,.
\label{int-pq}
\end{equation}
The gamma function has a pole at $D=3$.
To obtain $K_1$ in the limit  $D \to 3$, 
we need to expand the Feynman parameter integral in powers of $D-3$:
\begin{eqnarray}
K_1 &=&  {\Gamma(3-D) \over 2 (4\pi)^D} \Lambda^{2(D-3)}
\left\{ \int_0^1 dx \int_0^{1-x} dy \,
        \left[ (1-x)(1-y) - {z^2 \over 4} \right]^{-3/2}               
\right.
\nonumber \\ 
&& \left. \hspace{-.5in}
\;-\; {D-3 \over 2} \int_0^1 dx \int_0^{1-x} dy \,
        \left[ (1-x)(1-y) - {z^2 \over 4} \right]^{-3/2}               
        \ln \left[ (1-x)(1-y) - {z^2 \over 4} \right]
\right\}\,,
\label{K1-int}
\end{eqnarray}
where $z = 1-x-y$.
The first integral in (\ref{K1-int}) can be computed analytically 
and has the value $4 \pi/3$.  The second integral has to be computed 
numerically and has the value $-9.43698$.
Extracting the pole in $D-3$ from the gamma function in (\ref{K1-int}) 
and keeping all terms that survive in the limit $D\to 3$, we obtain
\begin{equation}
K_1 \;=\; - {1 \over 96 \pi^2} \Lambda^{2(D-3)}
        \left[ {1 \over D-3} + 1.12646 + \gamma - \ln(4 \pi)
        \right]\,,
\label{K1-num}
\end{equation} 
where $\gamma$ is Euler's constant.
   
We next consider the integral $K_2$ in (\ref{K2-def}).
By introducing a Feynman parameter, it can be written
\begin{equation}
K_2 \;=\;  \int_0^1 dx \; (1-x)
\int_{\bf p} \int_{\bf q} 
\left( {1 \over [p^2 + xq^2 + x {\bf p} \cdot {\bf q} + \Lambda^2]^3}
\;-\;  {1 \over [(1-x)p^2 + xq^2 + \Lambda^2]^3} \right) .
\label{K2-fp}
\end{equation}
By shifting and rescaling the momentum variables, 
we can reduce the integral over the momenta to (\ref{int-pq}).
In the first term of (\ref{K2-fp}), we shift 
${\bf p} \to {\bf p} - {x \over 2} {\bf q}$ and then rescale 
${\bf q}$ by $\left( {x(4-x) \over 4} \right)^{-1/2}$.  
In the second term, we rescale ${\bf p}$ by $(1-x)^{-1/2}$
and ${\bf q}$ by $x^{-1/2}$.
After integrating over ${\bf p}$ and $\bf q$, we obtain 
\begin{equation}
K_2 \;=\; {\Gamma(3-D) \over 2 (4\pi)^D} \Lambda^{2(D-3)}
\int_0^1 dx\, (1-x) x^{-D/2}
\left[ (1-x/4)^{-D/2} - (1-x)^{-D/2} \right]    \,.  
\label{K2-x}
\end{equation}
To obtain $K_2$ in the limit  $D \to 3$, 
we need to expand the integrand in (\ref{K2-x}) in powers of $D-3$:
\begin{eqnarray}
K_2 &=& {\Gamma(3-D) \over 2 (4\pi)^D} \Lambda^{2(D-3)}
\left\{ \int_0^1 dx 
\left( {8(1-x) \over \sqrt{x^3(4-x)^3}} 
        - {1 \over \sqrt{x^3(1-x)}}
        \right)
\right.
\nonumber \\ 
&& \left.  \hspace{-.5in}
\;-\; {D-3 \over 2} \int_0^1 dx
\left( {8(1-x) \over \sqrt{x^3(4-x)^3}} \ln {x(4-x) \over 4}
        - {1 \over \sqrt{x^3(1-x)}} \ln[x(1-x)]
        \right)
\right\}\,. 
\label{K22}
\end{eqnarray}
The integrals can be evaluated analytically.
Extracting the pole in $D-3$ from the gamma function and keeping all terms
that survive in the limit $D\to 3$, we obtain
\begin{equation}
K_2 \;=\; {\sqrt{3} \over 64 \pi^3} \Lambda^{2(D-3)}
         \left[ {1\over D-3} + {4 \pi \over 3 \sqrt{3}}
         - {1 \over 2} \ln {3 \over 4} - 1  + \gamma - \ln(4\pi)
         \right]\,.
\end{equation} 

\subsection{Two-loop diagrams for the Free Energy Density}
\label{twoloopdia}

The two-loop vacuum diagrams that contribute to
$-i\Omega_2(\mu,\overline{v}_0)$ are shown in 
Fig.~\ref{twoloop}.
These diagrams can be reduced to momentum integrals by
integrating over the loop energies using the identities in
Appendix~\ref{energyint}. Expressed in terms of the integrals 
$I_{m,n}$ and $J_{l,m,n}$ defined in (\ref{imn}) and (\ref{jlmn}),
the contributions of the individual diagrams to
$\Omega_2(\mu,\overline{v}_0)$ are
\begin{eqnarray}
      \Omega_2^{(a)} & = & {3 \over 64} g\, I_{1,1}^2 
\\
      \Omega_2^{(b)} & = & {1 \over 32} g\, I_{-1,-1} I_{1,1}
\\
      \Omega_2^{(c)} & = & {3 \over 64} g\, I_{-1,-1}^2 
\\
      \Omega_2^{(d)} & = & 
         - {3 \over 16} m g^2 \overline{v}_0^2 J_{1, 1, 1} 
\\
      \Omega_2^{(e)} & = & 
         {3 \over 8} m g^2 \overline{v}_0^2 J_{0, 0, 1} 
\\
      \Omega_2^{(f)} & = & 
         - {1 \over 16} m g^2 \overline{v}_0^2 J_{-1, -1, 1} 
\\
      \Omega_2^{(g)} & = & 
         - {1 \over 8} m g^2 \overline{v}_0^2 J_{-1,0, 0}  \, .
\end{eqnarray}
Adding up these diagrams, we get~(\ref{ome2b}).

\newpage
\section*{Figure Captions}

\begin{enumerate}

\item\label{feynman}
   Propagators and interaction vertices 
   for the real-valued fields $\xi$ and $\eta$.

\item\label{twoloop} 
   Two-loop vacuum diagrams that
   contribute to the thermodynamical potential.

\item\label{feynpsi} 
  Propagator and interaction vertices for the 
  complex-valued field $\psi$.

\item\label{scat:22} 
Diagrams for $2 \to 2$ scattering: 
        the tree-level diagram~(a), a one-loop diagram~(b), 
        and a two-loop diagram~(c).
   
\item\label{scat:33} 
One-particle-reducible diagrams for $3 \to 3$ scattering: 
        a tree-level diagram~(a) and a one-loop diagram~(b).

\item\label{scat:33-1PI} 
One-particle-irreducible (1PI) diagrams for $3 \to 3$ scattering: 
        a  one-loop diagram~(a) 
        and two-loop diagrams~(b), (c), and (d).

\end{enumerate}

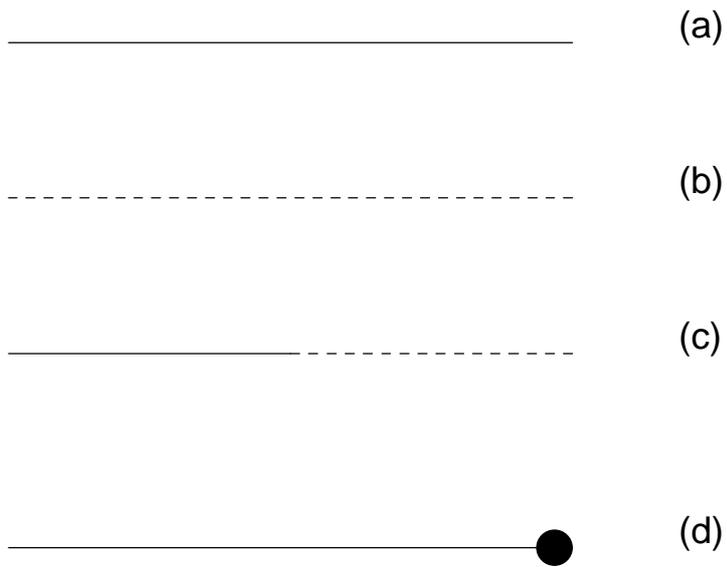
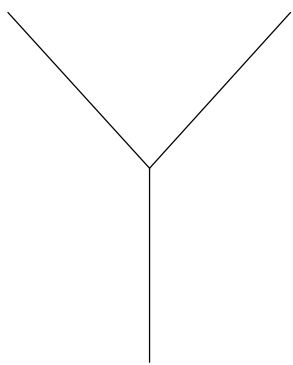
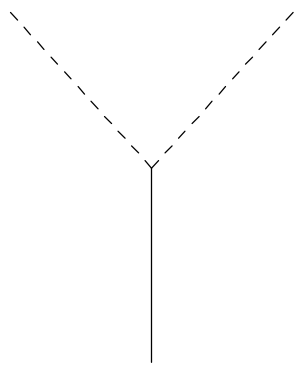
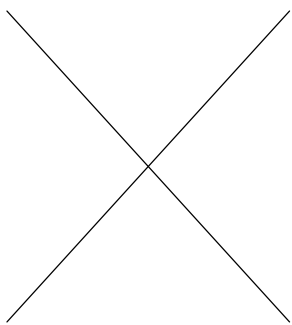
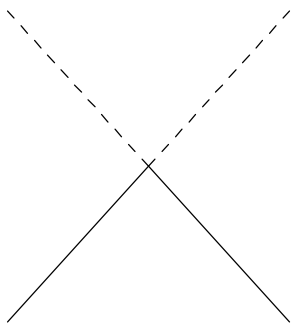
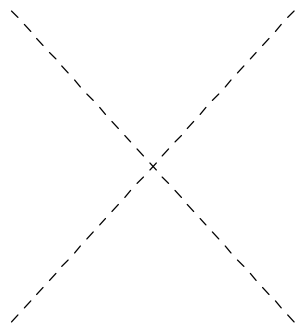
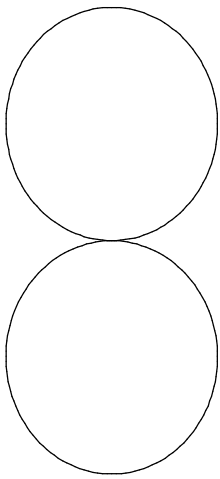
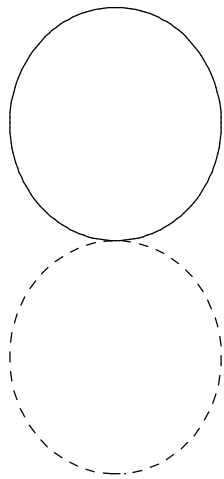
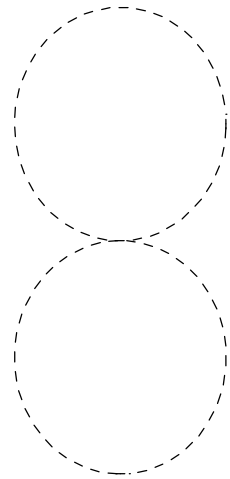
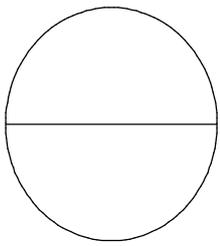
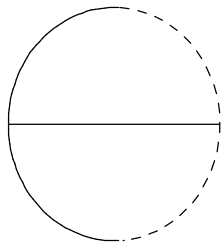
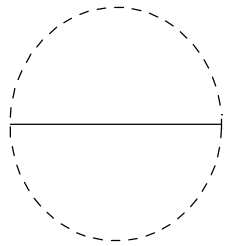
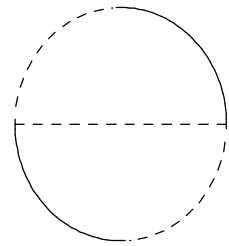
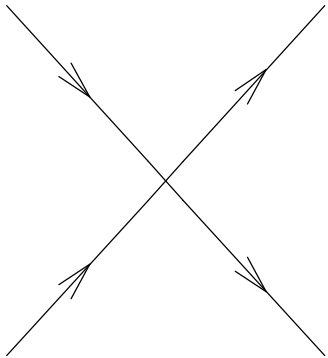
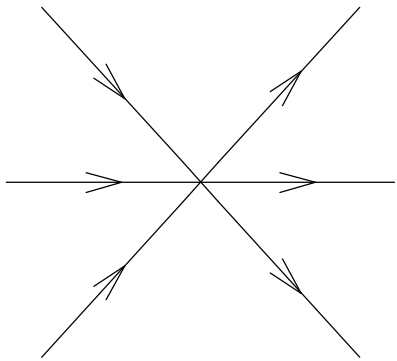
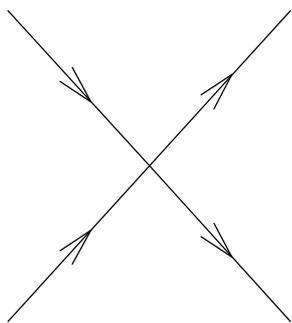
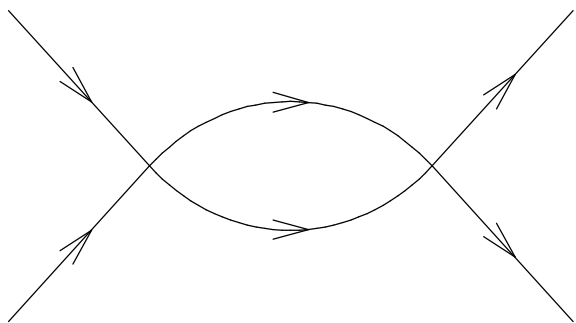
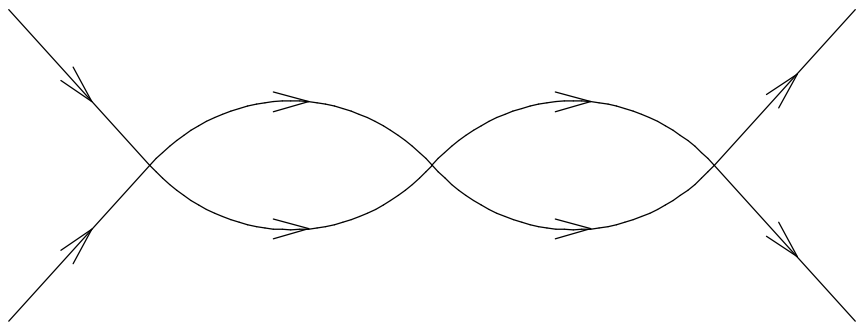
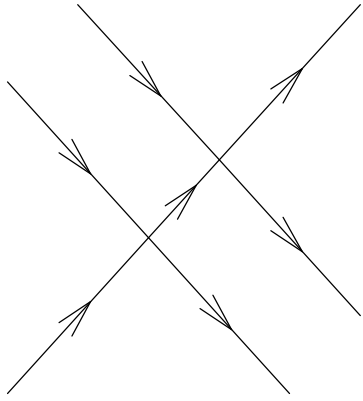
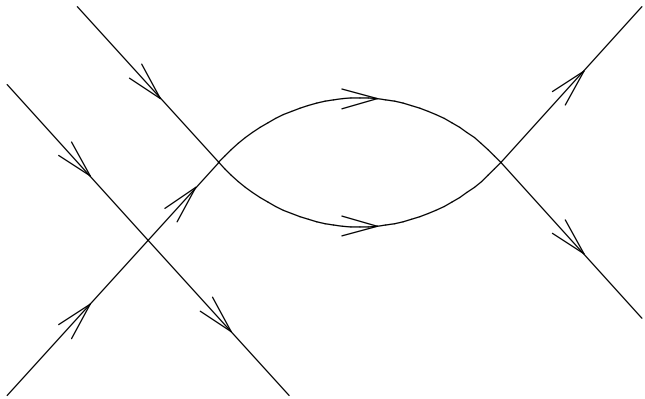
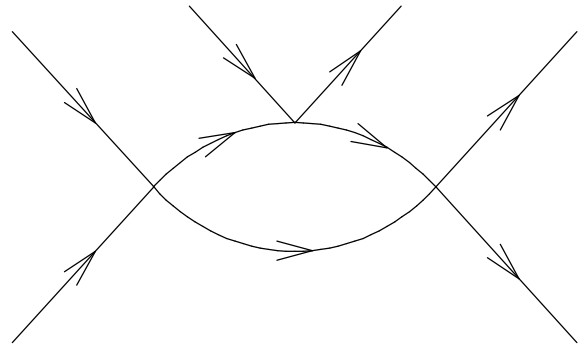
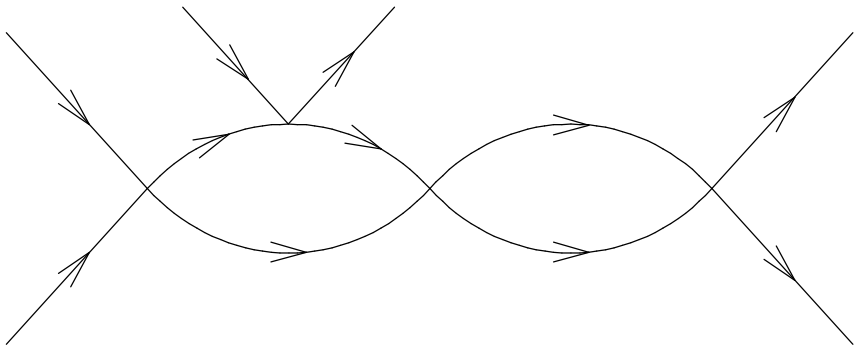
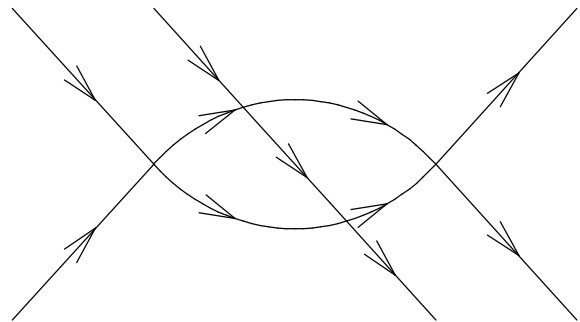
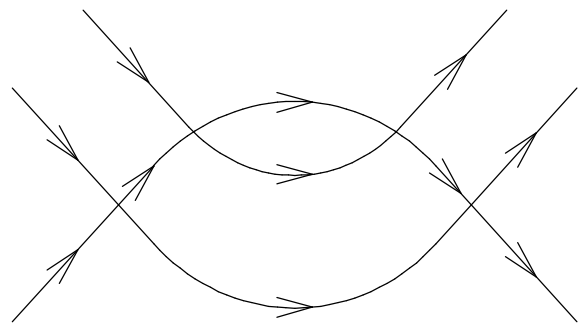
\end{document}